\documentclass[pre,aps,10pt,twocolumn,noshowpacs,amsmath,amssymb,superscriptaddress]{revtex4-1}

\usepackage{graphicx}
\usepackage{amsfonts}
\usepackage{epsfig}
\usepackage{epstopdf}
\usepackage{accents}
\usepackage{mathtools}

\usepackage{color}

\usepackage[pdfencoding=auto,psdextra]{hyperref}

\DeclareMathOperator{\re}{Re}
\DeclareMathOperator{\im}{Im}

\DeclareMathOperator{\Tr}{Tr}

\begin{document}

\title{Local impedance on a rough surface of a chiral $p$-wave superconductor}

\author{S.~V.~Bakurskiy}
\affiliation{Skobeltsyn Institute of Nuclear Physics, Lomonosov Moscow State University, Moscow 119991, Russia}
\affiliation{Moscow Institute of Physics and Technology, 141700 Dolgoprudny, Russia}

\author{Ya.~V.~Fominov}
\email[Corresponding author. E-mail address: ]{yfominov@hse.ru}
\affiliation{L.~D.\ Landau Institute for Theoretical Physics RAS, 142432 Chernogolovka, Russia}
%\affiliation{Moscow Institute of Physics and Technology, 141700 Dolgoprudny, Russia}
\affiliation{National Research University Higher School of Economics, 101000 Moscow, Russia}

\author{A.~F.~Shevchun}
\affiliation{Institute of Solid State Physics, 142432 Chernogolovka, Russia}
\affiliation{Moscow Institute of Physics and Technology, 141700 Dolgoprudny, Russia}

\author{Y.~Asano}
\affiliation{Department of Applied Physics and Center for Topological Science and Technology, Hokkaido University, Sapporo 060-8628, Japan}
\affiliation{Moscow Institute of Physics and Technology, 141700 Dolgoprudny, Russia}

\author{Y.~Tanaka}
\affiliation{Department of Applied Physics, Nagoya University, Nagoya 464-8603, Japan}
\affiliation{Moscow Institute of Physics and Technology, 141700 Dolgoprudny, Russia}

\author{M.~Yu.~Kupriyanov}
\affiliation{Skobeltsyn Institute of Nuclear Physics, Lomonosov Moscow State University, Moscow 119991, Russia}
\affiliation{Moscow Institute of Physics and Technology, 141700 Dolgoprudny, Russia}

\author{A.~A.~Golubov}
\affiliation{Moscow Institute of Physics and Technology, 141700 Dolgoprudny, Russia}
\affiliation{Faculty of Science and Technology and MESA+ Institute for Nanotechnology, University of Twente, 7500 AE Enschede, The Netherlands}

\author{M.~R.~Trunin}
\affiliation{National Research University Higher School of Economics, 101000 Moscow, Russia}
\affiliation{Institute of Solid State Physics, 142432 Chernogolovka, Russia}

\author{H.~Kashiwaya}
\affiliation{National Institute of Advanced Industrial Science and Technology (AIST), Tsukuba 305-8568, Japan}

\author{S.~Kashiwaya}
\affiliation{Department of Applied Physics, Nagoya University, Nagoya 464-8603, Japan}

\author{Y.~Maeno}
\affiliation{Department of Physics, Kyoto University, Kyoto 606-8502, Japan}

\date{31 October 2018}

\begin{abstract}
We develop a self-consistent approach for calculating the local impedance at a rough surface of a chiral $p$-wave superconductor. Using the quasiclassical Eilenberger-Larkin-Ovchinnikov formalism, we numerically find the pair potential, pairing functions, and the surface density of states taking into account diffusive electronic scattering at the surface.
The obtained solutions are then employed for studying the local complex conductivity and surface impedance in the broad range of microwave frequencies (ranging from subgap to above-gap values).
We identify anomalous features of the surface impedance caused by generation of odd-frequency superconductivity at the surface. The results are compared with experimental data for Sr$_2$RuO$_4$ and provide a microscopic explanation of the phenomenological two-fluid model suggested earlier to explain anomalous features of the microwave response in this material.
\end{abstract}

\maketitle

\tableofcontents

\section{Introduction}
\label{Intro}

Studying the surface properties of unconventional and topological superconductors is one of the central topics in modern solid-state physics \cite{Buchholtz1981,Hara1986,Hu1994,Tanaka1995,Kashiwaya2000,Yakovenko2004}. An important case is superconductivity in Sr$_2$RuO$_4$ with possible chiral $p$-wave spin-triplet pairing and nontrivial surface properties, which remains a subject of intensive experimental and theoretical investigation for a long time \cite{Maeno,Luke,Ishida,Mackenzie,Asano,Kikugawa,Kikugawa2,Nelson,MS,Kashiwaya11,Kittaka,Liu,Kallin,Maeno2016,Mackenzie17}. A number of theoretical works investigate formation of surface Andreev bound states and possible spontaneous surface currents (due to broken time-reversal symmetry) in $p$-wave superconductors and in superfluid $^3$He (which is a charge-neutral realization of $p$-wave superconductivity) \cite{Zhang,Matsumoto99,Tanaka2007a,Tanaka6,Higashitani2,Nagato2011,TanakaRev,Higashitani,Matsumoto2013,Bakurskiy1,LuBo,Suzuki,Bakurskiy2,
Nagato2018,Etter2018,Miyawaki2018}.

Measuring the microwave response may provide important information for understanding surface properties of unconventional superconductors \cite{Gough,Ormeno,Baker}. However, a fully consistent microscopic approach for calculating the surface impedance in these materials is not yet formulated. In recent experimental studies of the properties of Sr$_2$RuO$_4$ in microwave cavities, the results of measurements were interpreted in terms of the phenomenological two-fluid model \cite{Gough,Ormeno,Baker}. On the other hand, existing theoretical approaches \cite{Glazman,Kharitonov} for calculating the surface impedance are not applicable for clean anisotropic materials. A step forward was done in Ref.~\cite{Asano-Fominov}, where it was shown that the main anomalous contribution to the impedance is provided by a rough surface layer, in which the superconducting pair potential becomes isotropic. The approach developed in Refs.~\cite{Glazman,Kharitonov} can then be applied to calculation of the response functions.

In this work, we extend the results of Ref.~\cite{Asano-Fominov} by developing a self-consistent approach to the problem of calculation of the surface impedance in a chiral $p$-wave superconductor. We calculate the pair potential, pairing functions, and the density of states at the rough surface in a chiral $p$-wave superconductor. The obtained microscopic characteristics are then applied to studying the local complex conductivity and the surface impedance. The results for the surface impedance are compared with experiments performed on Sr$_2$RuO$_4$ samples and provide microscopic explanation of the phenomenological two-fluid model with finite quasiparticle fraction at zero temperature, introduced earlier \cite{Gough,Ormeno,Baker} to explain anomalies in this material.

The paper is organized as follows.
In Sec.~\ref{Sec1}, we introduce our theoretical approach and discuss the obtained results.
In Sec.~\ref{Sec2}, we present experimental results for the surface impedance of Sr$_2$RuO$_4$.
Relation between the theoretical and experimental results is discussed in Sec.~\ref{Sec3}.
Our results are summarized in Sec.~\ref{sec:conclusions}.
Details of theoretical derivation are presented in the Appendices.

Throughout the paper, we employ the units with $\hbar = k_B = 1$.

\section{Theory}
\label{Sec1}

\subsection{Method}
\label{subsec:method}

We consider a chiral $p$-wave superconductor, occupying the half-space $x<0$, with a flat surface producing diffusive scattering of quasiparticles (this is the limit of the ``rough'' surface, as opposed to the specularly reflecting one).

We describe the system in the framework of the quasiclassical
Eilenberger-Larkin-Ovchinnikov equations \cite{Eilenberger,LO1968}. The bulk $p$-wave
superconductor is assumed to satisfy the clean limit conditions (infinite scattering time), then the equations take the following form \cite{comment_fnotations} (see Appendix~\ref{app:derivation}):
\begin{gather}
2\omega_n f +v\cos\theta \frac{d f}{dx}= -2i \Delta g,
\label{El0} \\
2\omega_n f^{+} -v\cos\theta \frac{d f^{+}}{dx}= 2i \Delta^* g,
\label{El0a} \\
v\cos\theta \frac{d g}{dx} = i\Delta^* f + i\Delta f^{+} .
\label{El0b}
\end{gather}
The normal [$g(x,\theta,\omega_n)$] and anomalous [$f(x,\theta,\omega_n)$ and $f^{+}(x,\theta,\omega_n)$] Eilenberger-Larkin-Ovchinnikov functions satisfy the normalization condition
\begin{equation} \label{normal0}
g^2+f f^+ =1.
\end{equation}
Other notations have the following meaning: $\theta$ is the angle between the $x$ axis (normal to the surface) and the direction of electron Fermi velocity $v$, and
integer $n$ enumerates the Matsubara frequencies $\omega_n =\pi T(2n+1)$ with $T$ being temperature. The pair potential $\Delta(x,\theta)$ in a chiral $p$-wave superconductor ($p_x+i p_y$ state) can be decomposed into two components, $\Delta =\Delta_{x} (x) \cos \theta +i\Delta_{y}(x) \sin\theta$.

\begin{figure}[t]
\begin{minipage}[t]{0.04\columnwidth}
\vspace{2mm}
(a)
\end{minipage}%
\begin{minipage}[t]{0.96\columnwidth}
\vspace{0mm}
\includegraphics[width=0.96\columnwidth]{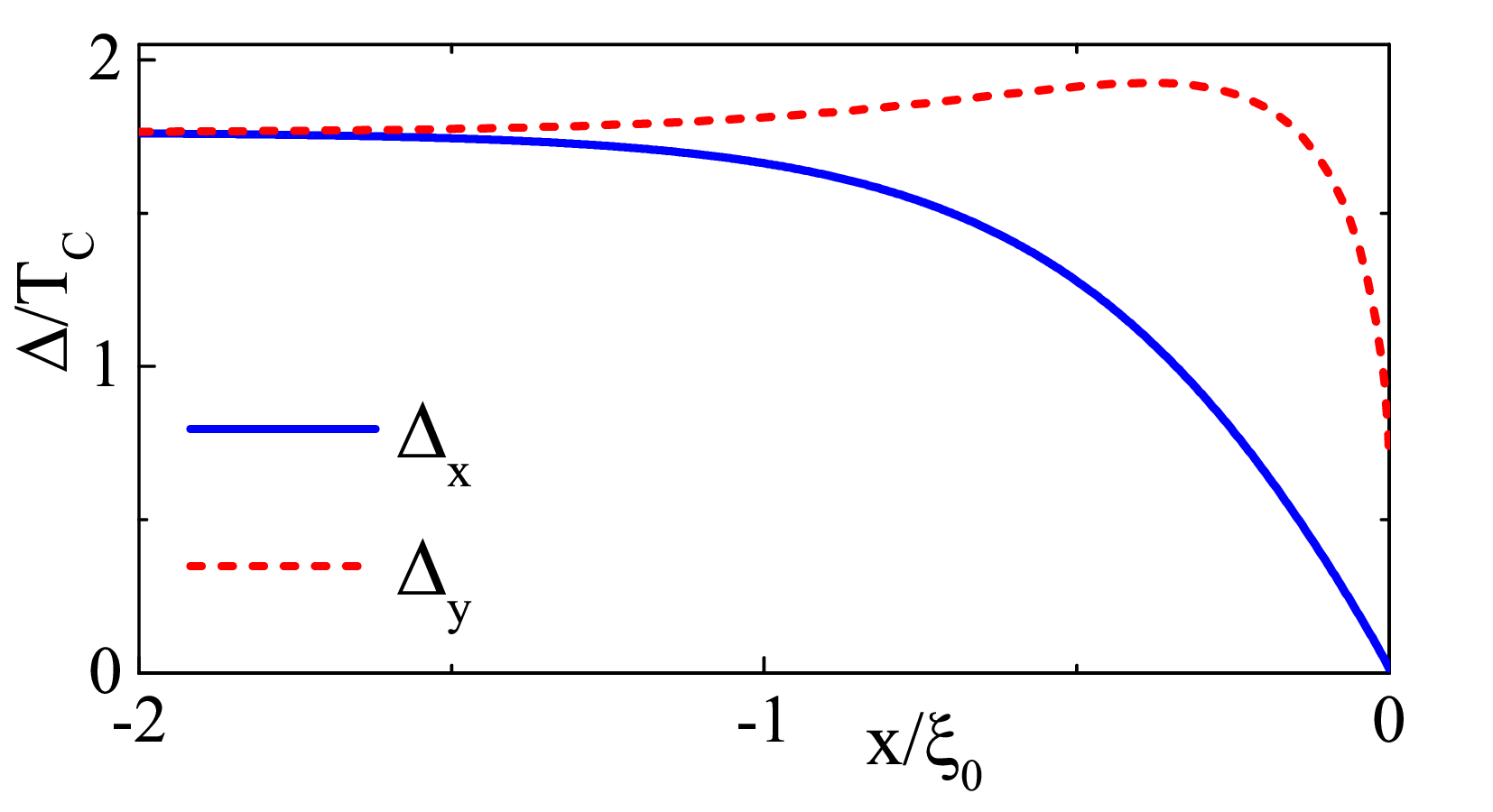}
\end{minipage}
\vspace{2mm}
\begin{minipage}[t]{0.04\columnwidth}
\vspace{2mm}
(b)
\end{minipage}%
\begin{minipage}[t]{0.96\columnwidth}
\vspace{0mm}
\includegraphics[width=0.96\linewidth]{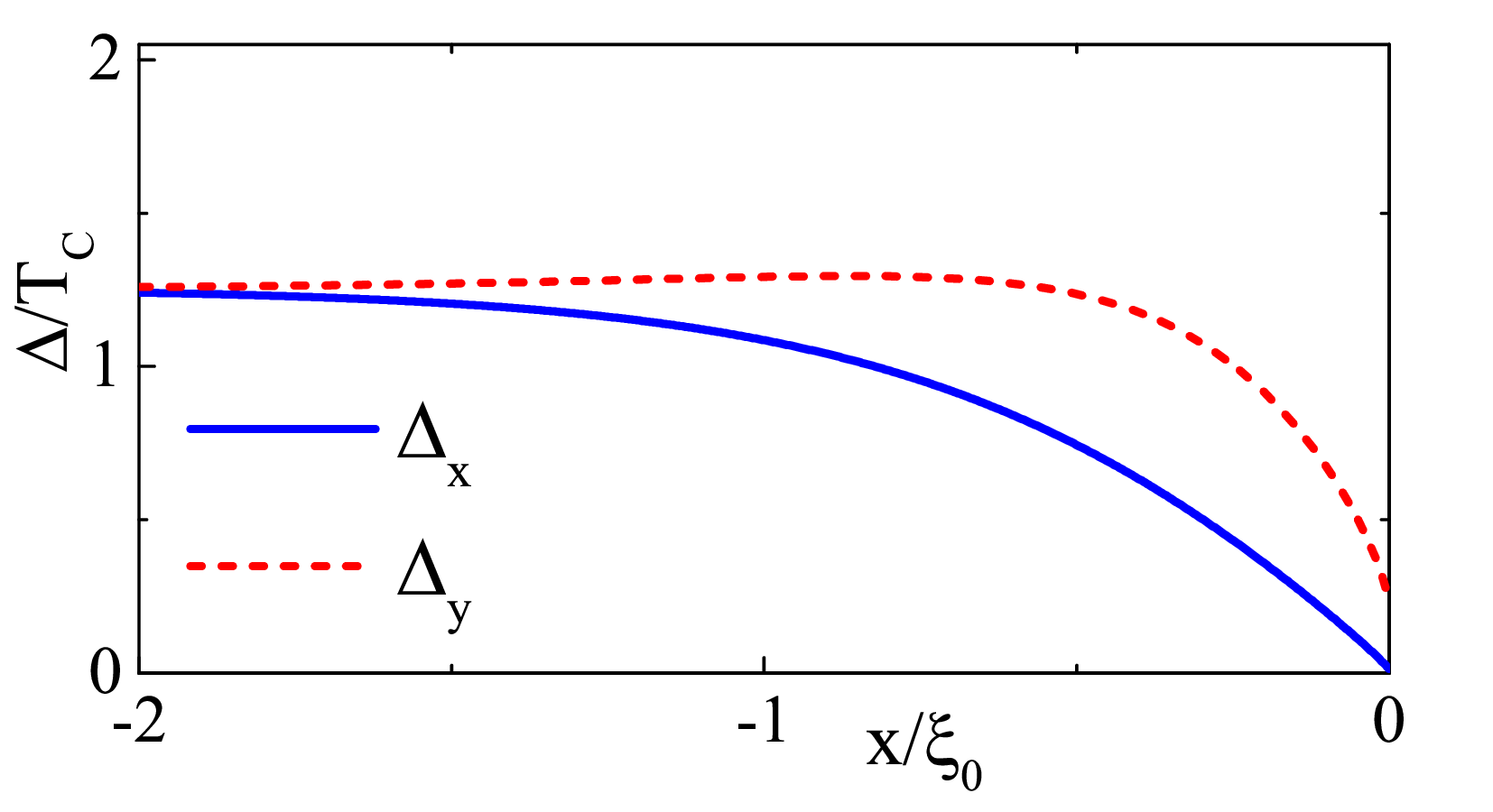}
\end{minipage}
\caption{Spatial distribution of the $\Delta_x$ and $\Delta_y$ components of the pair potential in the vicinity of the rough surface at (a)~$T=0.2 T_c$ and (b)~$T=0.8 T_c$. Our calculations imply an extremely rough limit such that reflected quasiparticles have the uniform angular distribution for any incidence angle.
As a result, incident trajectories are anisotropic, while reflected ones are isotropic, and after angular integration in the self-consistency equation we find small but finite $\Delta_x(0)$ [in contrast to the case of a specular interface that would lead to $\Delta_x(0)=0$ due to symmetry of the chiral $p$-wave state] \cite{Bakurskiy1,Bakurskiy2}.
}
\label{Del1}
\end{figure}

\begin{figure*}[t]
\begin{minipage}[t]{0.04\columnwidth}
\vspace{6mm}
(a)
\end{minipage}%
\begin{minipage}[t]{0.96\columnwidth}
\vspace{0mm}
\includegraphics[width=0.96\columnwidth]{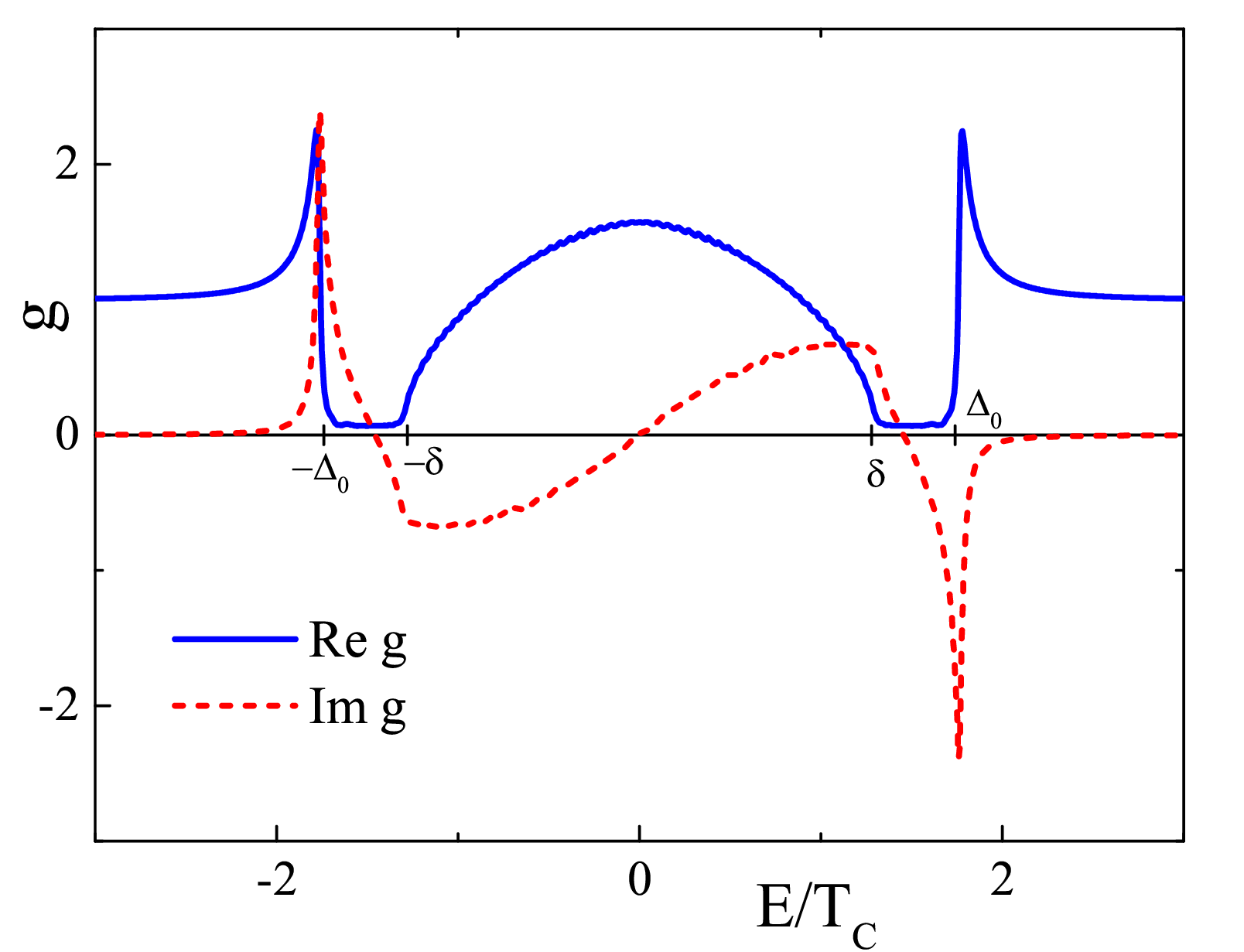}
\end{minipage}
\hfill
\begin{minipage}[t]{0.04\columnwidth}
\vspace{6mm}
(b)
\end{minipage}%
\begin{minipage}[t]{0.96\columnwidth}
\vspace{0mm}
\includegraphics[width=0.96\columnwidth]{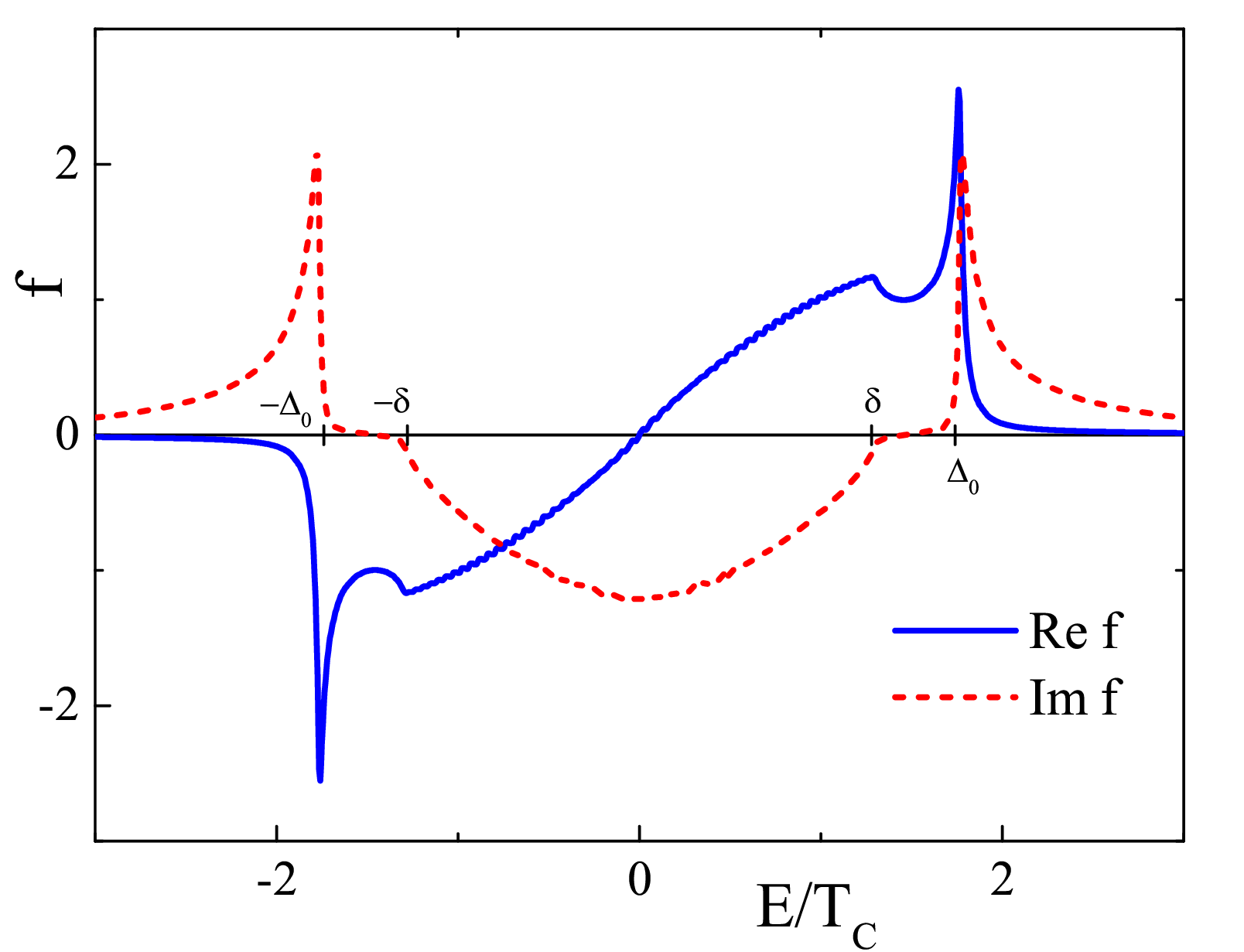}
\end{minipage}
\caption{Real and imaginary components of the Green functions (a)~$g(E)$ and (b)~$f(E)$ calculated at the outer surface of the structure at $T=0.2 T_c$. The DOS (normalized to the normal-metallic value) is given by $\re g$ and demonstrates two distinct energy intervals determined by the surface Andreev band ($|E|<\delta$) and the bulk band ($|E|>\Delta_0$). This classification is also relevant for other components as can be seen from the figures. Note that $\re g$ and $\im f$ should vanish inside the band gap (at $\delta < |E| < \Delta_0$), while their nonzero values in the figures are a numerical artifact originating from taking a small but finite imaginary part of energy ($E+i0$) in order to achieve convergence of our numerical procedure for calculating the retarded Green functions.}
\label{FG func}
\end{figure*}

\begin{figure}[t]
\center{\includegraphics[width=\columnwidth]{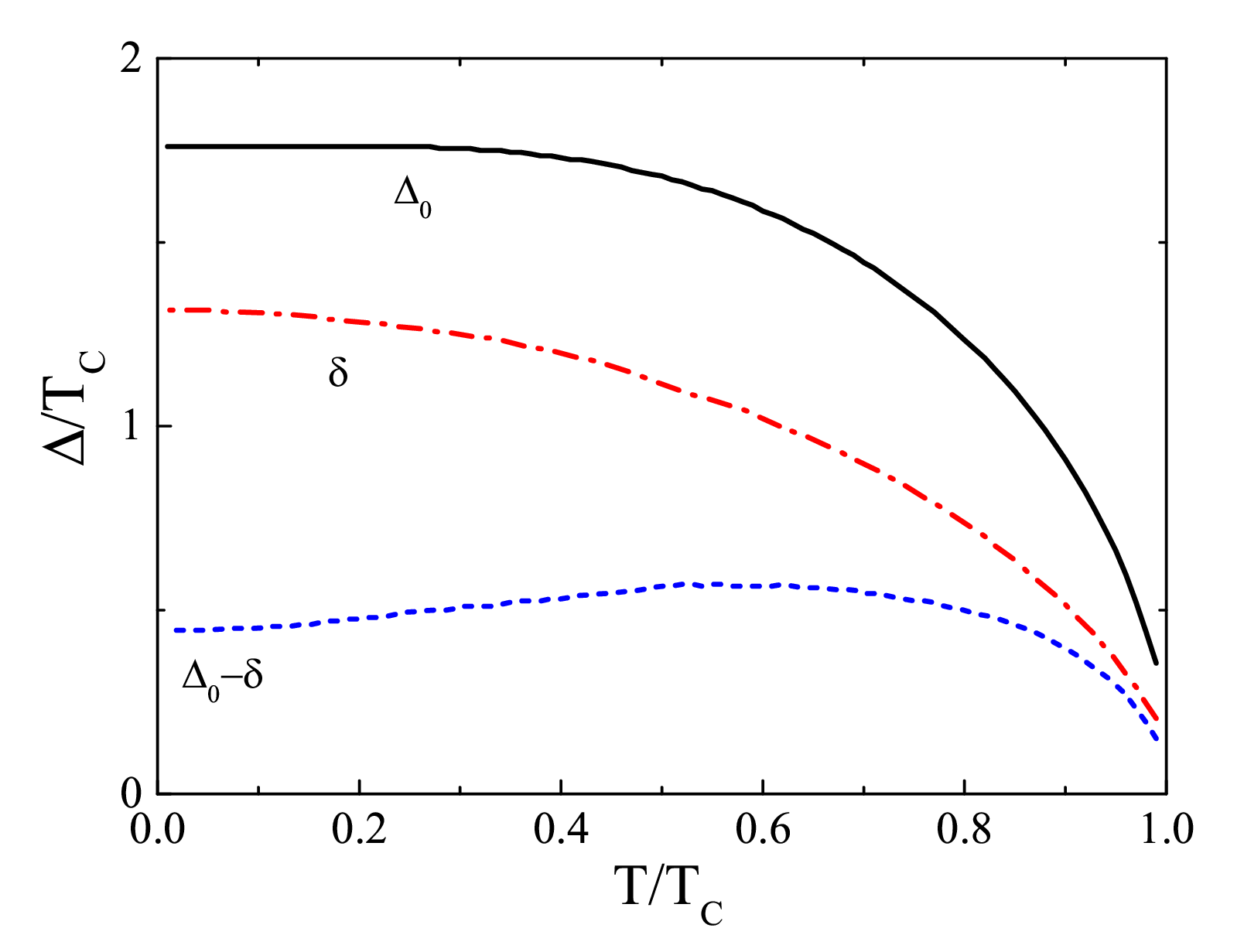}}
\caption{Temperature dependencies of the characteristic energies: half-width of the bulk gap $\Delta_0$ (solid line), half-width of the surface Andreev band $\delta$ (dashed line), and width of the surface band gap $\Delta_0-\delta$ (short-dashed line).}
\label{DelT}
\end{figure}

Equations (\ref{El0})-(\ref{El0b}) should be supplemented with the
self-consistency equation, which can be written (see Appendix~\ref{app:self-cons}) in the form
\begin{align}
\Delta _{x}\ln \frac{T}{T_{c}}+2\pi T\sum_{\omega_n>0} \left[ \frac{\Delta _{x}}{\omega_n} +\left\langle 2 \cos\theta \im f(\theta) \right\rangle \right] &= 0,
\label{Sc0b} \\
\Delta _{y}\ln \frac{T}{T_{c}}+2\pi T\sum_{\omega_n>0} \left[ \frac{\Delta _{y}}{\omega_n} -\left\langle 2 \sin\theta \re f(\theta) \right\rangle \right] &= 0.
\label{Sc0c}
\end{align}
Here, the angle brackets denote the angular averaging, $\left\langle ...\right\rangle =(1/2\pi )\int_{0}^{2\pi }(...)d\theta$,
and $T_{c}$ is the critical temperature of the superconductor.

Properties of rough superconductor surfaces can be described
\cite{Bakurskiy1,Bakurskiy2} in the framework of the Ovchinnikov model \cite{Ovchinnikov}. Within this model, the surface roughness is modelled by a thin diffusive normal-metal layer covering the superconductor. The thickness $d$ of the normal layer is assumed to be such that $l\ll d \ll \xi_0$ (where $l = v\tau$ is the mean free path and $\xi_0 = v/2\pi T_c$ is the coherence length).
Inside the layer (occupying the area $0 < x < d$), the Eilenberger-Larkin-Ovchinnikov equations \cite{Eilenberger,LO1968} (see Appendix~\ref{app:derivation}) transform to
\begin{align}
v\cos\theta \frac{d f}{dx} &= \frac{\left\langle f\right\rangle g -\left\langle g\right\rangle f}{\tau},
\label{El2a} \\
v\cos\theta \frac{d f^{+}}{dx} &= -\frac{\left\langle f^{+}\right\rangle g -\left\langle g\right\rangle f^{+}}{\tau},
\label{El2b} \\
v\cos\theta \frac{d g}{dx} &= \frac{\left\langle f^{+}\right\rangle f -\left\langle f\right\rangle f^{+}}{2\tau}.
\label{El2c}
\end{align}
Here, the $\omega_n$ terms, which could be written similarly to Eqs.\ (\ref{El0}) and (\ref{El0a}), are neglected due to a large value of $1/\tau$. Physically, the limit of very rough interface that we consider means that reflected quasiparticles have the uniform angular distribution for any incidence angle.

The Green functions should be continuous along each quasiclassical trajectory.
The resulting set of equations (\ref{El0})-(\ref{El2c}) can be solved numerically following the procedure developed in Refs.\ \cite{Bakurskiy1,Bakurskiy2}. In the course of this procedure, Eqs.\ (\ref{El0})-(\ref{El2c}) are treated as a system of linear equations, while $\Delta$, $\left\langle f\right\rangle $, $\left\langle f^{+}\right\rangle$, and
$\left\langle g\right\rangle$ are determined self-consistently.

The results of the self-consistent solution for the pair potential are illustrated in Fig.~\ref{Del1} in the two cases of relatively low [panel (a)] and high [panel (b)] temperatures.
In the bulk ($x \to -\infty$), the two components of the pair potential are equal, $\Delta_x=\Delta_y$. However, their behavior is different near the surface and depends on properties of the surface in accordance with previous results \cite{Higashitani2,Bakurskiy1}.
While mirror reflection would completely suppress $\Delta_x$ at the surface due to symmetry reasons, we observe that roughness leads to a small but finite value (the smallness is numerical, not parametric).
At the same time, $\Delta_y$ is not suppressed by a mirror surface, while roughness leads to a noticeable suppression of this component [still, this suppression is not so strong as in the case of $\Delta_x$; as a result, $\Delta_y(0)$ is much larger than $\Delta_x(0)$].

The spatial scales of suppression in the cases of $\Delta_x(x)$ and $\Delta_y(x)$ differ only by a numerical factor (which does not contain any large or small parameter).
The characteristic scale along each trajectory is $\xi_0$; however, the characteristic scale in the $x$ direction depends on the trajectory angle [this can be seen from the $\cos\theta$ factor multiplying the Fermi velocity in Eqs.\ (\ref{El0})-(\ref{El0b})]. Since the $\Delta_x$ component in our geometry is mainly contributed by quasiparticles with normal incidence ($\cos\theta$ close to 1), the corresponding suppression length is just $\xi_0$, see Fig.~\ref{Del1}. On the other hand, $\Delta_y$ is mainly determined by quasiparticles in the $p_y$ lobe, moving almost parallel to the surface (small $\cos\theta$). The integral effect on $\Delta_y$ is such that this component is suppressed on a length scale several times smaller than $\xi_0$ \cite{Bakurskiy2}, see Fig.~\ref{Del1}.

As a result, in the vicinity of the surface, $\Delta_y(x)$ is larger than $\Delta_x(x)$, which means that the pair potential is predominantly $p_y$ wave. The absolute value of the $\Delta_y$ component in this situation has a tendency to the bulk value of the pair potential in purely $p_y$-wave superconducting state, which exceeds the bulk value of $\Delta_y$ in the chiral $p_x+ i p_y$ pair potential \cite{comment:chiralvspy}.
This phenomenon leads to a maximum of $\Delta_y(x)$ at a distance from the surface smaller than $\xi_0$. The maximum is more pronounced
at lower temperatures, see Fig.~\ref{Del1}(a).

Having calculated the pair potential $\Delta (x,\theta)$ at different temperatures with the help of the Matsubara representation,
we switch to the representation of real energy $E$ and find the (retarded) Green
functions $f(E)$, $f^+(E)$, and $g(E)$. To this end, we substitute $\omega_n=- i (E+i0)$ and numerically solve the boundary problem defined by Eqs.\ (\ref{El0})-(\ref{El0b}) and (\ref{El2a})-(\ref{El2c}).

\subsection{Results for the Green functions}
\label{subsec:GF}

Below, we discuss properties of the Green functions calculated at the outer surface of the structure
(i.e., at $x=d$ although $d$ is very small in our calculations, $d\ll \xi_0$ \cite{comment:x=0}).
These Green functions are isotropicized due to strong impurity scattering inside the normal layer. The results of calculations are demonstrated in Fig.~\ref{FG func}, where we plot $\re g(E)$, $\im g(E)$, $\re f(E)$, and $\im f(E)$.

The real part of the normal Green function, $\re g(E)$, represents the surface density of states (DOS) normalized to the normal-metallic value. The DOS is an even function of $E$. The peaks of the surface DOS at $|E|=\Delta_0$ are inherited from the coherence peaks in the bulk superconductor (note that the chiral $p$-wave state is fully gapped with isotropic gap equal to $\Delta_0$).
They appear due to effective proximity effect between the bulk superconductor and the diffusive surface layer (that models the rough surface).
More generally, we can say that the whole continuum of the states at $|E|>\Delta_0$ (the bulk band) in the surface DOS originates from the bulk quasiparticle states of the superconductor.

At the same time, the DOS below $\Delta_0$ demonstrates a wide band of surface Andreev bound states \cite{Aoki2005,Higashitani2}. This is the surface effect, not present in the bulk, and related to reflections of quasiparticles from the surface, which changes the pair potential felt by quasiparticles (due to the anisotropic nature of superconductivity in our system).
The surface Andreev band is dispersive, as a consequence of the chiral $p_x+i p_y$ symmetry of the pair potential with internal superconducting phase difference between different trajectories (this contrasts the $p_x$ case with a sharp Andreev surface peak at zero energy).
The amplitude of the DOS in the vicinity of zero energy is significant and even exceeds the normal-metallic DOS in the limit of rough surface, see Fig.~\ref{FG func}.
This becomes obvious from the normalization condition (\ref{normal0}), which at $E=0$, due to vanishing of $\im g(0)$ and $\re f(0)$, reduces to
\begin{equation} \label{normaliz}
\left[ \re g(0) \right]^2 = 1 + \left[ \im f(0) \right]^2,
\end{equation}
which is clearly larger than unity.
We denote the half-width of the Andreev band by $\delta$, which is smaller than $\Delta_0$ (the half-width of the bulk superconducting gap).
Finally, the interval of energies between $\delta$ and $\Delta_0$ can be called the effective band gap (between the Andreev band and the bulk band).

The two energy scales, $\delta$ and $\Delta_0$, turn out to be characteristic energies not only for the DOS but for all the components of the Green functions, see Fig.~\ref{FG func}. Both these energies are suppressed by temperature (obviously vanishing at $T_c$), while their difference (the band gap) is slightly nonmonotonic with a shallow maximum, see Fig.~\ref{DelT}.
The real part of $g$ and the imaginary part of $f$ are even functions of energy $E$, while the imaginary part of $g$ and the real part of $f$ are odd (the symmetry depends on the choice of definitions for the Green functions and on the choice of the superconducting phase \cite{comment_fnotations}).

The quasiparticle states in the Andreev band can be interpreted \cite{Tanaka6} as a manifestation of the odd-frequency superconductivity \cite{Berezinskii1974,BVE_review} [enhanced DOS due to odd-frequency superconductivity, see Eq.\ (\ref{normaliz}), was pointed out in Refs.\ \cite{Tanaka2007a,Yokoyama2008}]. The odd-frequency property is clearly pronounced as the symmetry of the anomalous Green function in the language of the Matsubara technique (odd dependence on the Matsubara frequency $\omega_n$), while in the real-energy ($E$) representation it relates retarded and advanced functions and therefore is not obvious from Fig.~\ref{FG func} (where we only plot retarded functions). Nevertheless, this unusual property will become evident later when we discuss the imaginary part of the conductivity.

Although the Andreev bound states at the surface of the chiral $p$-wave superconductor are also present in the case of specular surface, roughness ``emphasizes'' them. The surface itself breaks the rotational symmetry and leads to mixing of superconducting states with different angular momenta \cite{Tanaka2007b,Eschrig2007}. In its turn, disorder (surface roughness) suppresses all anisotropic harmonics, thus singling out the $s$-wave superconducting component, which inevitably possesses the odd-frequency symmetry in the triplet case (note that the bulk $p$-wave superconductivity is spin triplet) \cite{Tanaka2007a,Bakurskiy1}. The surface Andreev states that we find are a manifestation of this odd-frequency superconducting component.

The self-consistency for the pair potential [Eqs.\ (\ref{Sc0b}) and (\ref{Sc0c})] turns out to be important for the formation of a well-defined Andreev band of width $2\delta$, separated from the bulk band. This can be seen from comparison with Ref.~\cite{Asano-Fominov}, where similar calculations were performed for the case of rough surface but without taking into account the self-consistency. The obtained results also demonstrated enhanced subgap Andreev states but the Andreev band was merged with the bulk band without any band gap, so in this sense the Andreev band was not well defined.

Note that the Green functions at the free surface of the rough layer ($x=d$), which we discuss, are different from the functions at the interface between the clean superconductive and the rough layer ($x=0$) \cite{Nagato2011,Bakurskiy1}. The Green functions at the interface are not isotropic and have essentially different energy dependence inside the Andreev band ($|E| < \delta$). For example, the energy dependence of the DOS inside the Andreev band is nearly flat due to contributions from quasiparticle trajectories almost parallel to the interface ($\theta\approx \pi/2$) \cite{Nagato2011,Bakurskiy1}, in contrast to the dome-shaped DOS at the outer surface [Fig.~\ref{FG func}(a)].

\begin{figure*}[t]
\begin{minipage}[t]{0.04\columnwidth}
\vspace{6mm}
(a)
\end{minipage}%
\begin{minipage}[t]{0.96\columnwidth}
\vspace{0mm}
\includegraphics[width=0.96\columnwidth]{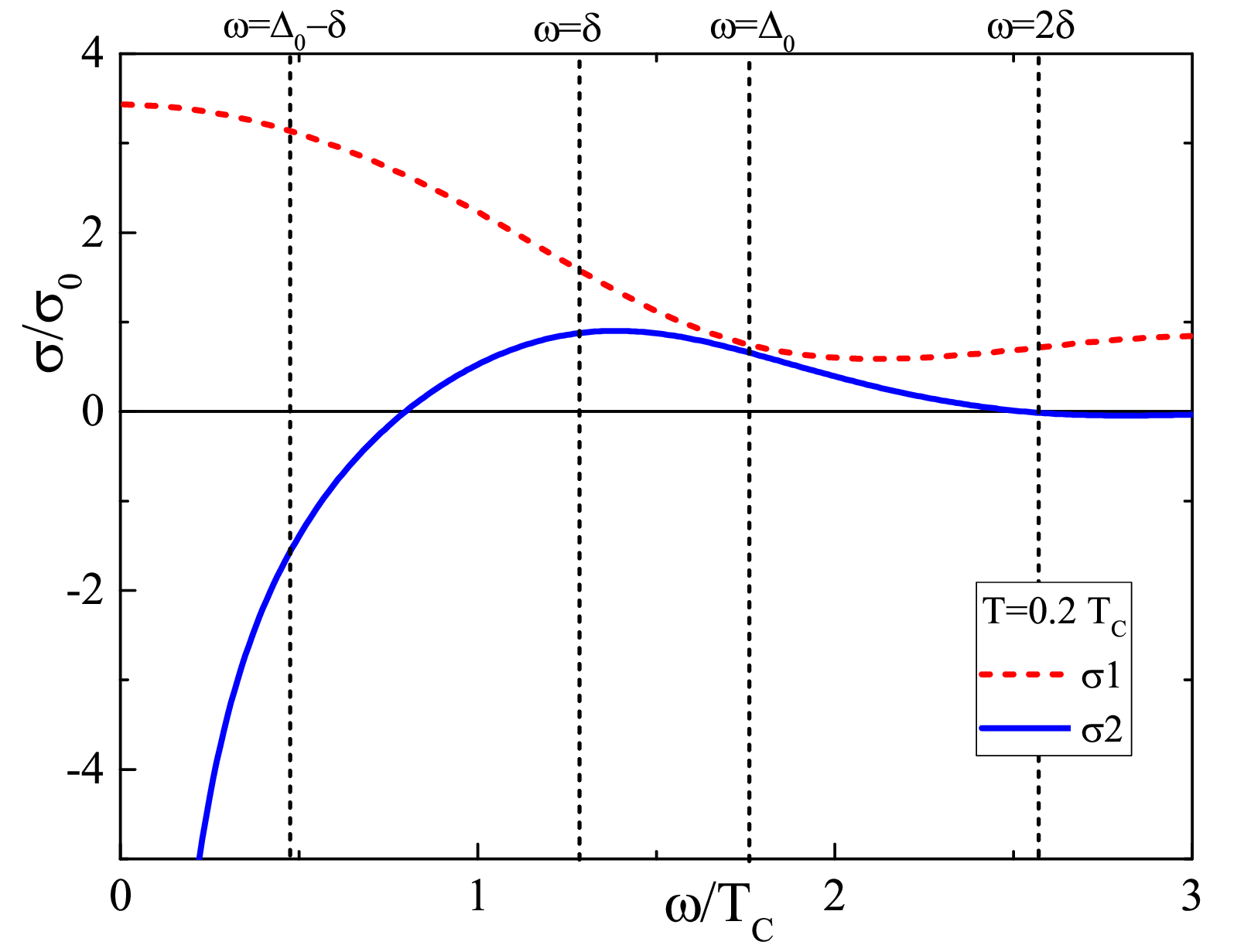}
\end{minipage}
\hfill
\begin{minipage}[t]{0.04\columnwidth}
\vspace{6mm}
(b)
\end{minipage}%
\begin{minipage}[t]{0.96\columnwidth}
\vspace{0mm}
\includegraphics[width=0.96\columnwidth]{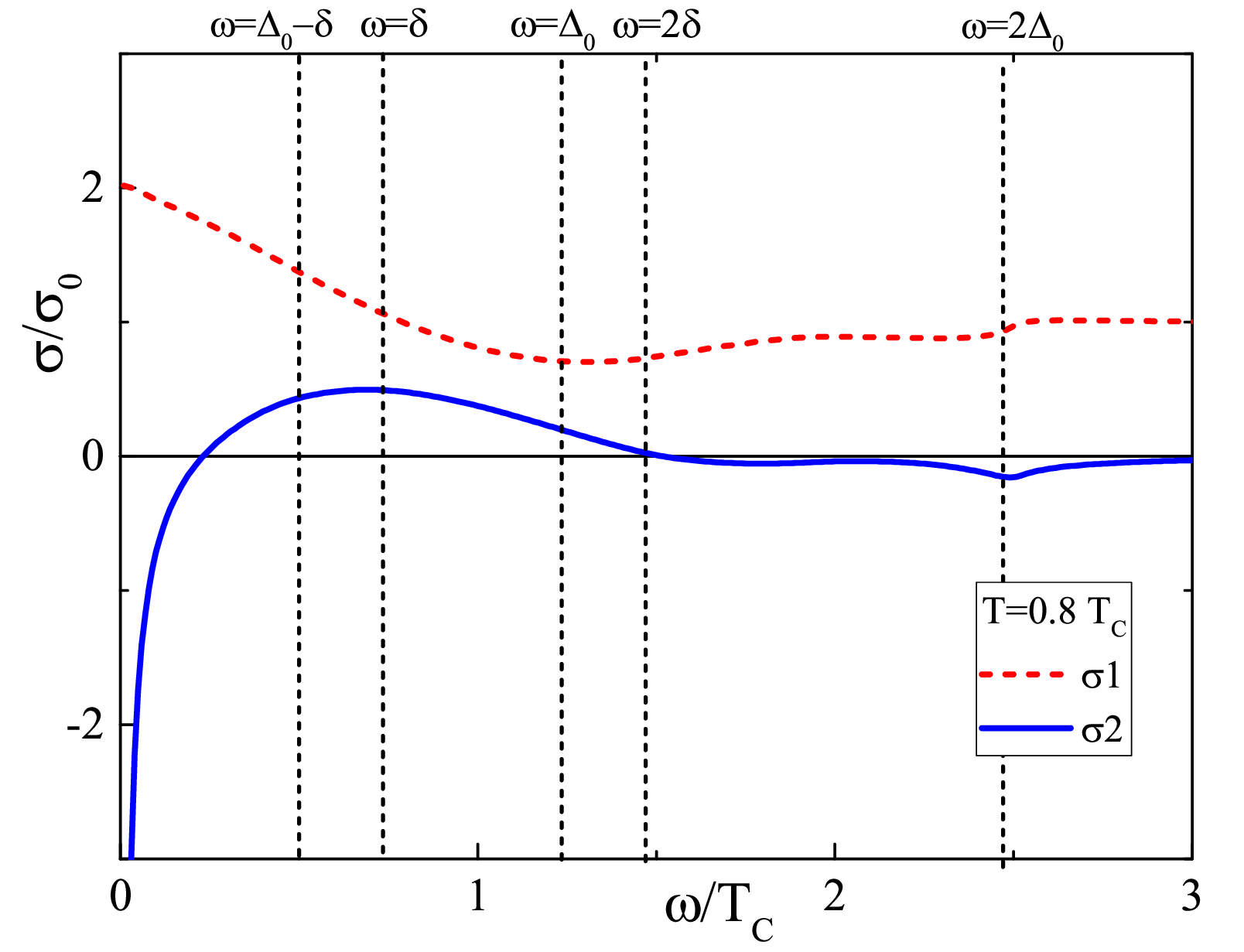}
\end{minipage}
\caption{Real $\sigma_1$ and imaginary $\sigma_2$ components of the conductivity at the rough surface of chiral $p_x+i p_y$ superconductor vs.\ frequency $\omega$ at (a)~low temperature $T=0.2 T_c$ and
(b)~high temperature $T=0.8 T_c$.  Vertical dashed lines demonstrate characteristic energies for the surface Green functions.}
\label{Sigma}
\end{figure*}

\begin{figure*}[t]
\begin{minipage}[t]{0.04\columnwidth}
\vspace{6mm}
(a)
\end{minipage}%
\begin{minipage}[t]{0.96\columnwidth}
\vspace{0mm}
\includegraphics[width=0.96\columnwidth]{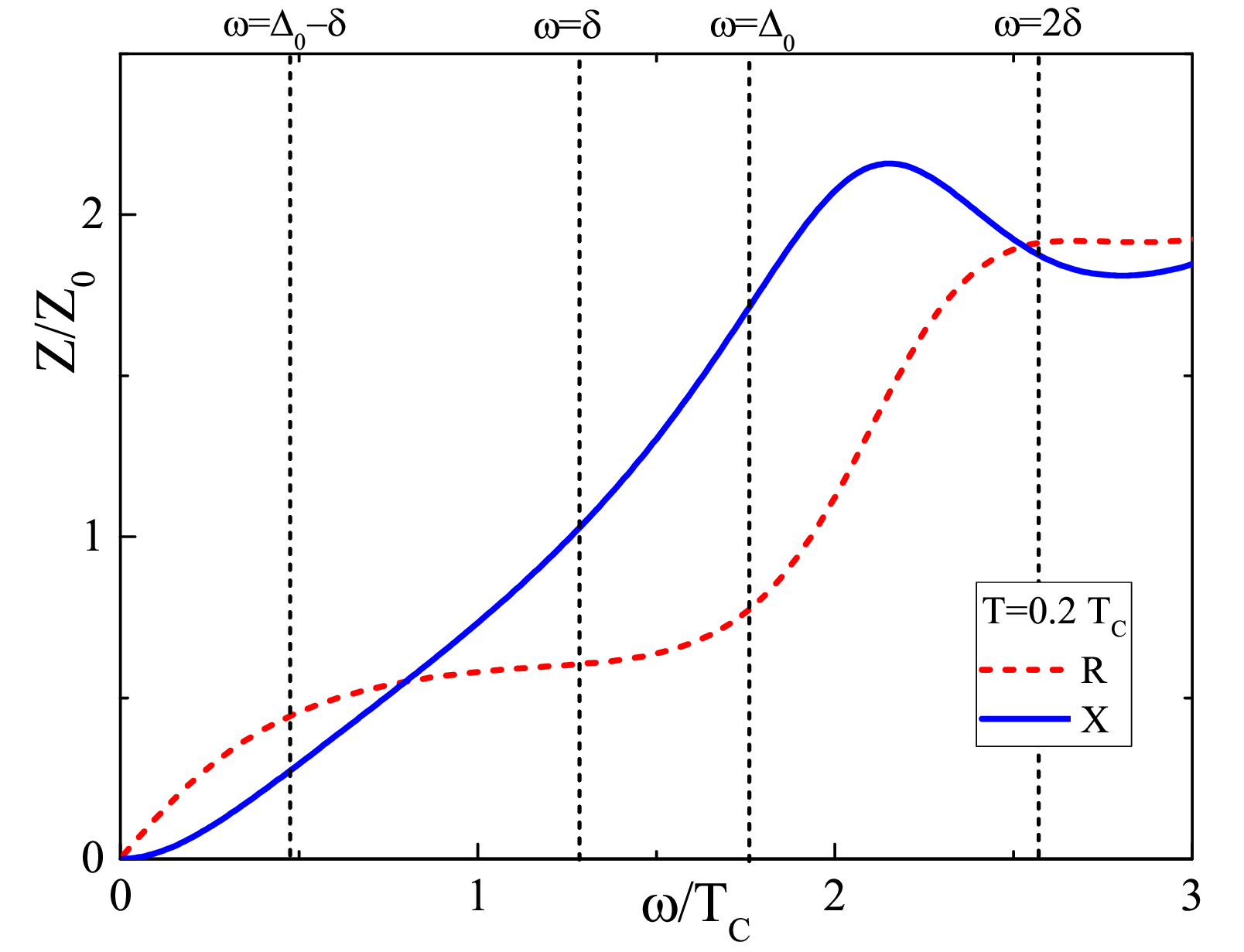}
\end{minipage}
\hfill
\begin{minipage}[t]{0.04\columnwidth}
\vspace{6mm}
(b)
\end{minipage}%
\begin{minipage}[t]{0.96\columnwidth}
\vspace{0mm}
\includegraphics[width=0.96\columnwidth]{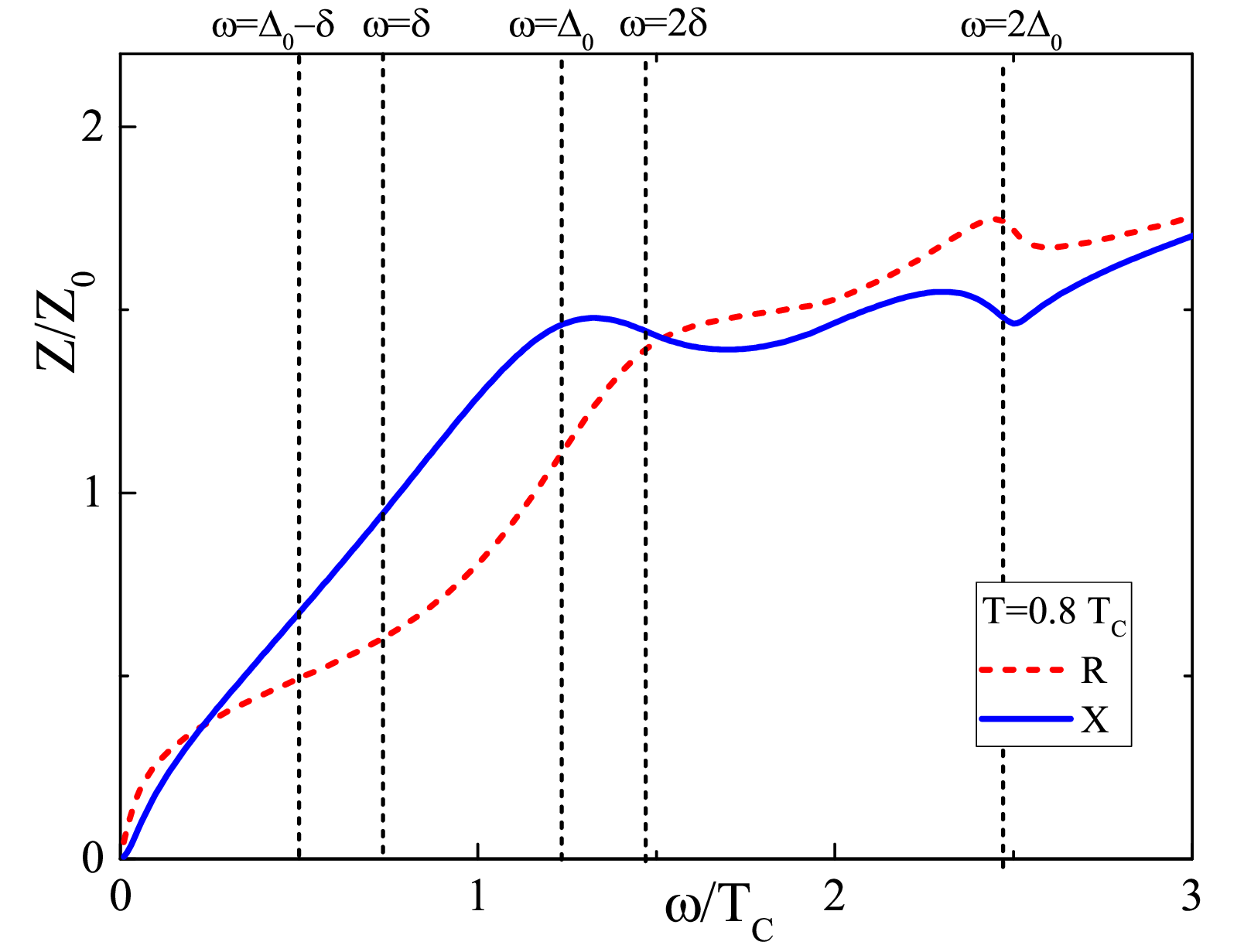}
\end{minipage}
\caption{Real $R$ and imaginary $X$ components of the local impedance $Z$ at the rough surface of chiral $p_x+i p_y$ superconductor vs.\ frequency $\omega$ at (a)~low temperature $T=0.2 T_c$ and
(b)~high temperature $T=0.8 T_c$.
The normalization factor $Z_0$ equals the normal-metallic $R$ [in the normal state, $R=X$ and we can write $Z(\omega) = (1-i)Z_0(\omega)$], taken at $\omega = T_c$.
Vertical dashed lines demonstrate characteristic energies for the surface Green functions.}
\label{Z_hw}
\end{figure*}

\subsection{Results for the complex conductivity}
\label{subsec:conductivity}

The Eilenberger Green functions at the outer surface of the structure are isotropic due to the rough (diffusive) surface layer and thus reduce to the Usadel Green functions. The complex conductivity $\sigma(\omega)$ at the surface can then be calculated with the help of the dirty-limit expression (\ref{sigma_all}) (where we substitute our effectively isotropic functions). Separating the real and imaginary parts, $\sigma(\omega) = \sigma_1(\omega)+i\sigma_2(\omega)$ and denoting
$E_\pm = E\pm \omega/2$,
we find
\begin{multline} \label{sigma1}
\frac{\sigma_1(\omega)}{\sigma_0} =
\frac{1}{2 \omega} \int\limits_{-\infty}^{\infty }
dE \left( \tanh \frac{E_+}{2T} - \tanh \frac{E_-}{2T} \right)
\\
\times \left[
\re g(E_+) \re g(E_-) + \im f(E_+) \im f(E_-)
\right],
\end{multline}
\vspace{-7mm}
\begin{multline} \label{sigma2b}
\frac{\sigma_2(\omega)}{\sigma_0} =
-\frac{1}{\omega} \int\limits_{-\infty}^{\infty }
dE \tanh \frac{E_-}{2T}
\\
\times \left[ \re g(E_-) \im g(E_+) - \im f(E_-) \re f(E_+) \right],
\end{multline}
where $\sigma_0$ is the Drude conductivity [note that in the normal metal, we reproduce the trivial dirty-limit result $\sigma_1(\omega)=\sigma_0$ and $\sigma_2(\omega)=0$].

We substitute into Eqs.\ (\ref{sigma1}) and (\ref{sigma2b}) the Green functions $g$ and $f$ calculated at the surface according to the procedure discussed above (see Secs.~\ref{subsec:method} and~\ref{subsec:GF}).
Conductivity $\sigma$ as a function of frequency $\omega$ is demonstrated in Fig.~\ref{Sigma} for low $T=0.2 T_c$ and high $T=0.8 T_c$ temperatures.

\subsubsection{Real part of conductivity}

The real part of conductivity, $\sigma_1(\omega)$, determines the dissipative response of quasiparticles and superconducting condensate to external electromagnetic field of frequency $\omega$. The dissipation occurs due to transitions between energies $E-\omega/2$ and $E+\omega/2$, as described by the products of the Green functions in Eq.\ (\ref{sigma1}). The difference of the hyperbolic tangents in the first line of Eq.\ (\ref{sigma1}) defines the energy window inside which the transitions contribute to dissipation. Qualitative behavior of $\sigma_1$ can be understood from considering the quasiparticle contribution (the $\re g \cdot \re g$ term), while the condensate contribution (the $\im f \cdot \im f$ term) only affects this behavior quantitatively. Note that Eq.\ (\ref{sigma1}) is written in a symmetrized form, in which the integrand is an even function of $E$.

The anomalous feature of the $\sigma_1(\omega)$ dependence in our system is the absence of threshold that usually characterizes this function at low temperatures in fully gapped superconductors. At the same time, despite the fully gapped nature of the chiral $p$-wave state in the bulk, the surface features the presence of the subgap Andreev band. Therefore, even at $T\to 0$, there are always surface states at arbitrarily low energy, which can participate in dissipation. Due to those states, the static dissipative conductivity $\sigma_1(0)$ is finite; moreover, it exceeds the normal-state value, see Fig.~\ref{Sigma}. The static limit is easily extracted from Eq.\ (\ref{sigma1}) (at any $T$):
\begin{equation} \label{sigma1(0)}
\frac{\sigma_1(0)}{\sigma_0} = \left[ \re g(0) \right]^2 + \left[ \im f(0) \right]^2 = 1 + 2 \left[ \im f(0) \right]^2.
\end{equation}
Figure~\ref{FG func} illustrates that the two terms in the right-hand side of the first equality (due to quasiparticles and due to the condensate, respectively) provide similar contributions, both exceeding 1. The last equality is written with the help of the normalization condition (\ref{normaliz}) and demonstrates that $\sigma_1(0) > \sigma_0$.

When $\omega$ increases, the anomalous intra-Andreev-band transitions provide smaller dissipation since not only the top of the dome but also energies closer to its sides are involved, and the latter correspond to smaller DOS. This explains the maximum of $\sigma_1$ at $\omega=0$.

Further increase of $\omega$ and/or increase of $T$ turn on two more types of dissipative processes: the Andreev-to-bulk-band and the conventional superconducting bulk-to-bulk-band transitions. The Andreev-to-bulk-band transitions appear at $\omega > \Delta_0-\delta$, i.e., when frequency is sufficient to overcome the band gap [note that while $\sigma_1$ is positive, the condensate contribution to $\sigma_1$ due to such transitions is negative, see Fig.~\ref{FG func}(b)]. The bulk-to-bulk-band transitions are effective at any $\omega$ if $T$ exceeds $\Delta_0$, so that there are many thermally excited quasiparticles above the bulk gap. On the other hand, at $T\ll \Delta_0$, the bulk-to-bulk-band transitions appear in the threshold manner when $\omega > 2\Delta_0$. In any case, when $\omega$ exceeds $2\Delta_0$, the dissipative conductivity tends to the normal-state Drude conductivity $\sigma_0$.

Note also that as frequency grows, the anomalous intra-Andreev-band transitions are turned off at $\omega > 2\delta$, since the energy jump corresponding to the transitions becomes larger than the width of the Andreev band.

Comparing the cases of (relatively) low and high temperatures, Figs.~\ref{Sigma}(a) and~\ref{Sigma}(b), we observe that the frequency scale at higher temperatures shrinks due to the decrease of the characteristic energies $\delta$ and $\Delta_0$ as functions of temperature.

We conclude that the subgap Andreev states qualitatively modify the $\sigma_1(\omega)$ behavior, leading to a low-frequency maximum at all temperatures.
This contrasts the conventional behavior in the gapped case, where $\sigma_1(\omega)$ is exponentially suppressed at low temperatures, $T\ll \Delta_0$.

\subsubsection{Imaginary part of conductivity}

The integral part of Eq.\ (\ref{sigma2b}) for $\sigma_2(\omega)$ is finite in the superconducting state in the limit $\omega\to 0$ [note that the integrand in Eq.\ (\ref{sigma2b}) becomes an even function of $E$ in this limit]. The imaginary part of conductivity is therefore inversely proportional to $\omega$ in this limit. The proportionality constant is standardly related \cite{Tinkham} to the density of superconducting electrons $n_s$ according to
\begin{equation} \label{sigma2}
\sigma_2(\omega) = \frac{n_s e^2}{m \omega},
\end{equation}
where $e$ and $m$ are electron's charge and mass, respectively. Equation (\ref{sigma2b}) yields
\begin{multline} \label{ns}
\frac{n_s(T)}n =
-\tau \int\limits_{-\infty}^{\infty }
dE \tanh \frac{E}{2T}
\\
\times \left[ \re g(E) \im g(E) - \im f(E) \re f(E) \right],
\end{multline}
where we normalize $n_s$ by the total density of electrons $n$.

Bulk superconductivity is characterized by positive $n_s$. As a manifestation of this fact, we see that
both terms in the square brackets in Eq.\ (\ref{ns}) give positive contribution to $n_s$ from the bulk band (i.e., from integration over energies $|E|>\Delta_0$), as follows from Fig.~\ref{FG func}.

At the same time, odd-frequency superconductivity is characterized by negative $n_s$. This implies an unconventional sign of the current response to external electromagnetic field, anomalous Meissner effect \cite{Higashitani1997,Walter1998,Tanaka2005,Yokoyama2011,Suzuki2014,Suzuki2015,Suzuki}, and anomalous behavior of the surface impedance \cite{Asano-Fominov}. In our calculations, we observe that both terms in the square brackets in Eq.\ (\ref{ns}) provide negative contribution to $n_s$ from the Andreev band (i.e., from integration over energies $|E|< \delta$), as follows from Fig.~\ref{FG func}. This is natural, since the surface Andreev states are a manifestation of the odd-frequency superconductivity \cite{Tanaka6,Tanaka2007b}.

As can be expected from Fig.~\ref{FG func}, the main contribution to $n_s$ is given by the Andreev band. The behavior of $\sigma(\omega)$ at small frequencies is therefore anomalous, since it corresponds to $n_s<0$. This is illustrated by Fig.~\ref{Sigma}. Comparison between the two different temperatures, shown in the figure, demonstrates that the absolute value of $n_s(T)$ is reduced with increasing temperature. This effect is mainly due to the $\tanh(E/2T)$ factor in Eq.\ (\ref{ns}), which suppresses the contribution from energies of the order of $T$ and smaller.

In the case of finite frequencies, the integrand in Eq.\ (\ref{sigma2b}) behaves in a complicated manner, changing its sign several times, and is therefore rather nontransparent from the point of view of qualitative understanding. The hyperbolic tangent does not lead to an energy window [like it was in the case of $\sigma_1(\omega)$], allowing ``transitions'' (products of the Green functions with arguments shifted by $\omega$) at all energies; only the energies corresponding to $|E_-| \lesssim T$ are effectively cut out. In addition to intra-Andreev-band and intra-bulk-band transitions, finite frequencies lead to the appearance of transitions between the Andreev and bulk bands.

At the same time, as $\omega$ grows, the anomalous Andreev-band contribution becomes suppressed, the conventional contributions take over, and $\sigma_2$ acquires the conventional (positive) sign. At $\omega \sim \delta$, we find a maximum of $\sigma_2(\omega)$.

From our previous discussion of characteristic energy scales, we can expect that the contribution to $\sigma_2(\omega)$ of the intra-Andreev-band processes should vanish when $\omega>2\delta$. Since this contribution is dominant at small $\omega$, the remaining $\sigma_2$ at $\omega>2\delta$ should be relatively small. This is indeed confirmed by Fig.~\ref{Sigma}. Interestingly, $\sigma_2(\omega)$ crosses zero in the vicinity of $\omega=2\delta$ (remaining small at larger frequencies), meaning that Andreev-to-bulk-band and bulk-to-bulk-band processes nearly compensate each other at this frequency. At larger frequencies, $\sigma_2(\omega)$ is negative but small; in this sense, this is a marginally anomalous regime.

We conclude that while subgap Andreev states preserve the $\sigma_2(\omega) \propto 1/\omega$ behavior at low frequencies,
the proportionality coefficient (containing $n_s$) becomes negative, in contrast to the standard gapped case in which it is positive.
The crossover to the high-frequency regime is then characterized by a positive maximum of $\sigma_2(\omega)$.

\begin{figure*}[t]
\begin{minipage}[t]{0.04\columnwidth}
\vspace{6mm}
(a)
\end{minipage}%
\begin{minipage}[t]{0.96\columnwidth}
\vspace{0mm}
\includegraphics[width=0.96\columnwidth]{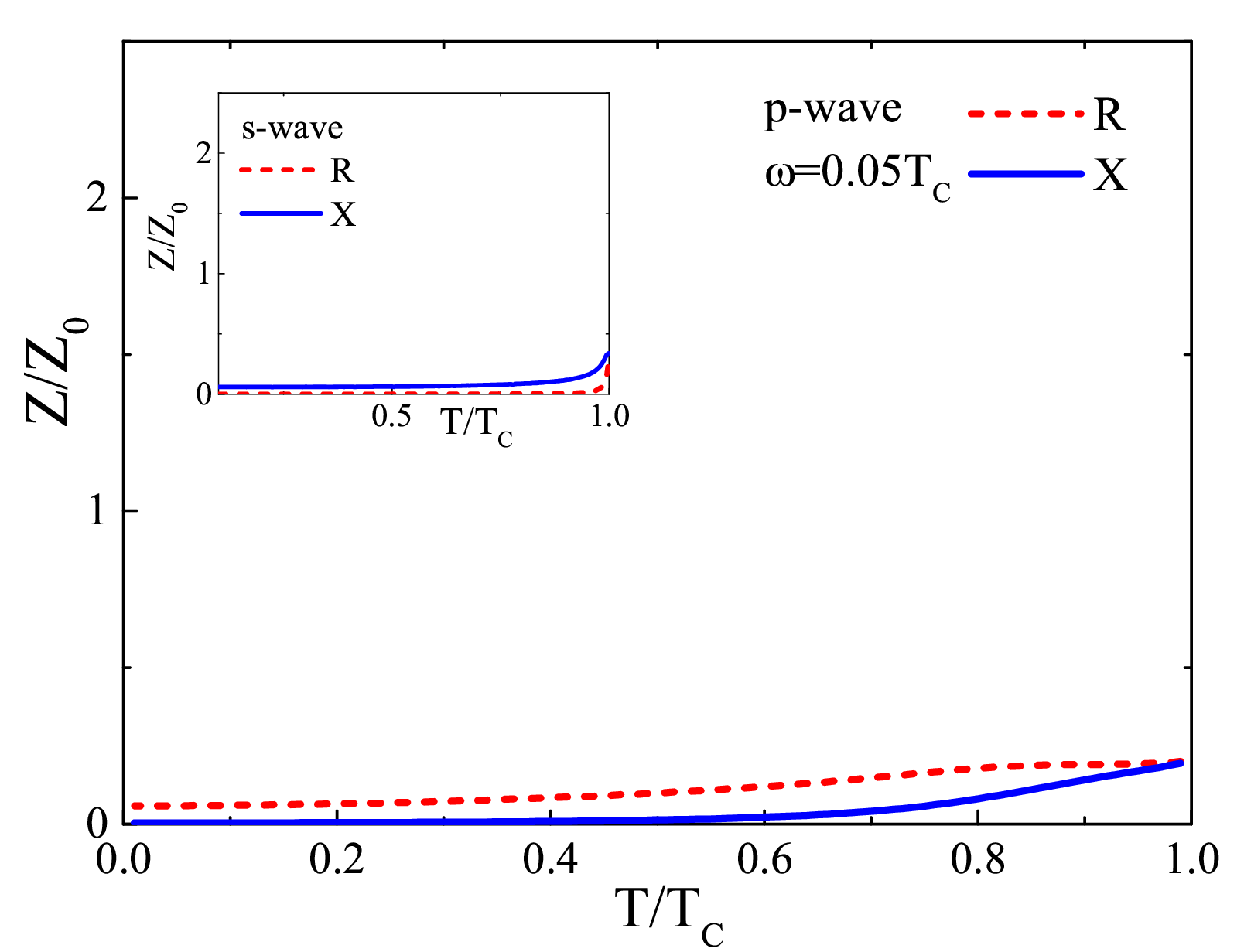}
\end{minipage}
\hfill
\begin{minipage}[t]{0.04\columnwidth}
\vspace{6mm}
(b)
\end{minipage}%
\begin{minipage}[t]{0.96\columnwidth}
\vspace{0mm}
\includegraphics[width=0.96\columnwidth]{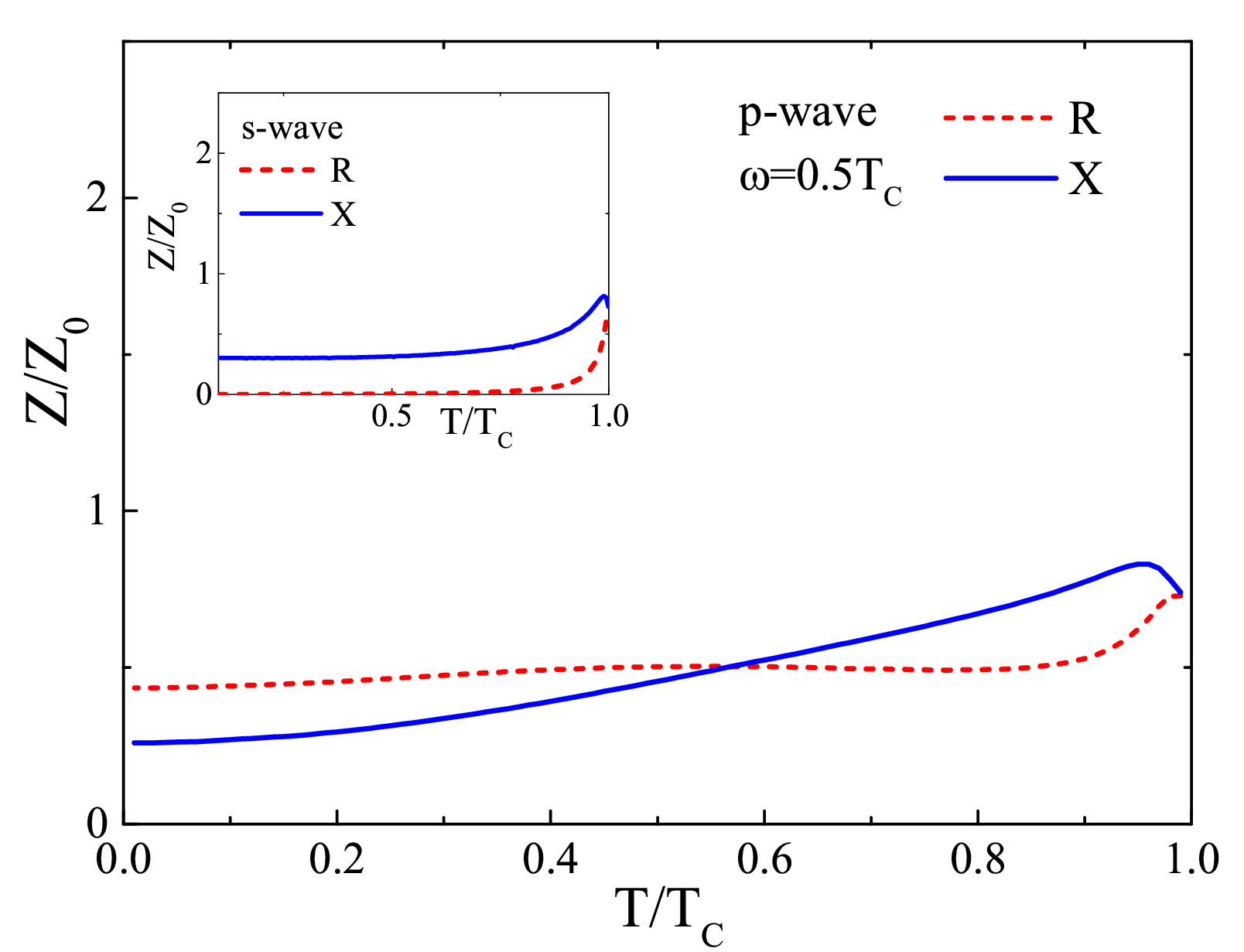}
\end{minipage} \\
\begin{minipage}[t]{0.04\columnwidth}
\vspace{6mm}
(c)
\end{minipage}%
\begin{minipage}[t]{0.96\columnwidth}
\vspace{0mm}
\includegraphics[width=0.96\columnwidth]{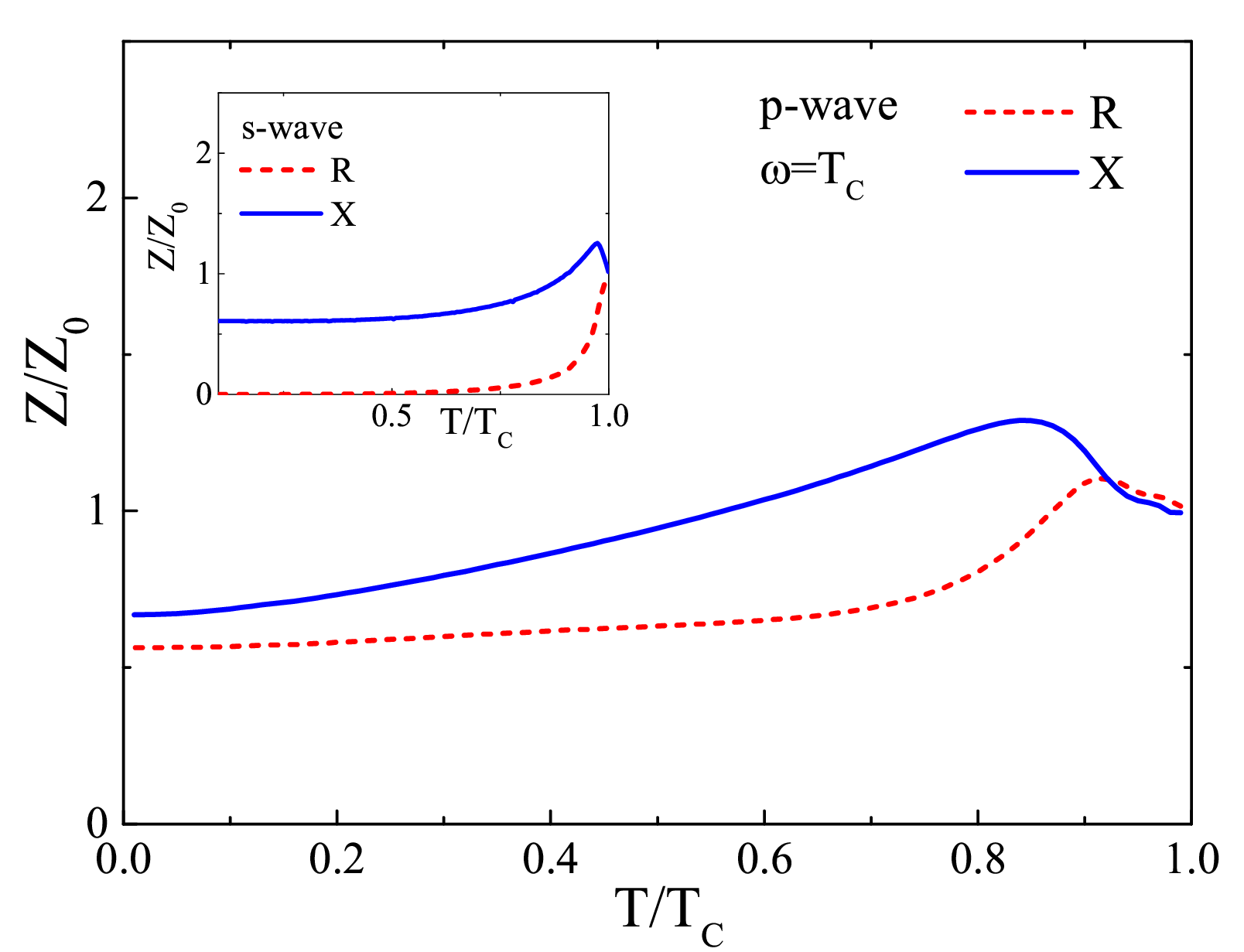}
\end{minipage}
\hfill
\begin{minipage}[t]{0.04\columnwidth}
\vspace{6mm}
(d)
\end{minipage}%
\begin{minipage}[t]{0.96\columnwidth}
\vspace{0mm}
\includegraphics[width=0.96\columnwidth]{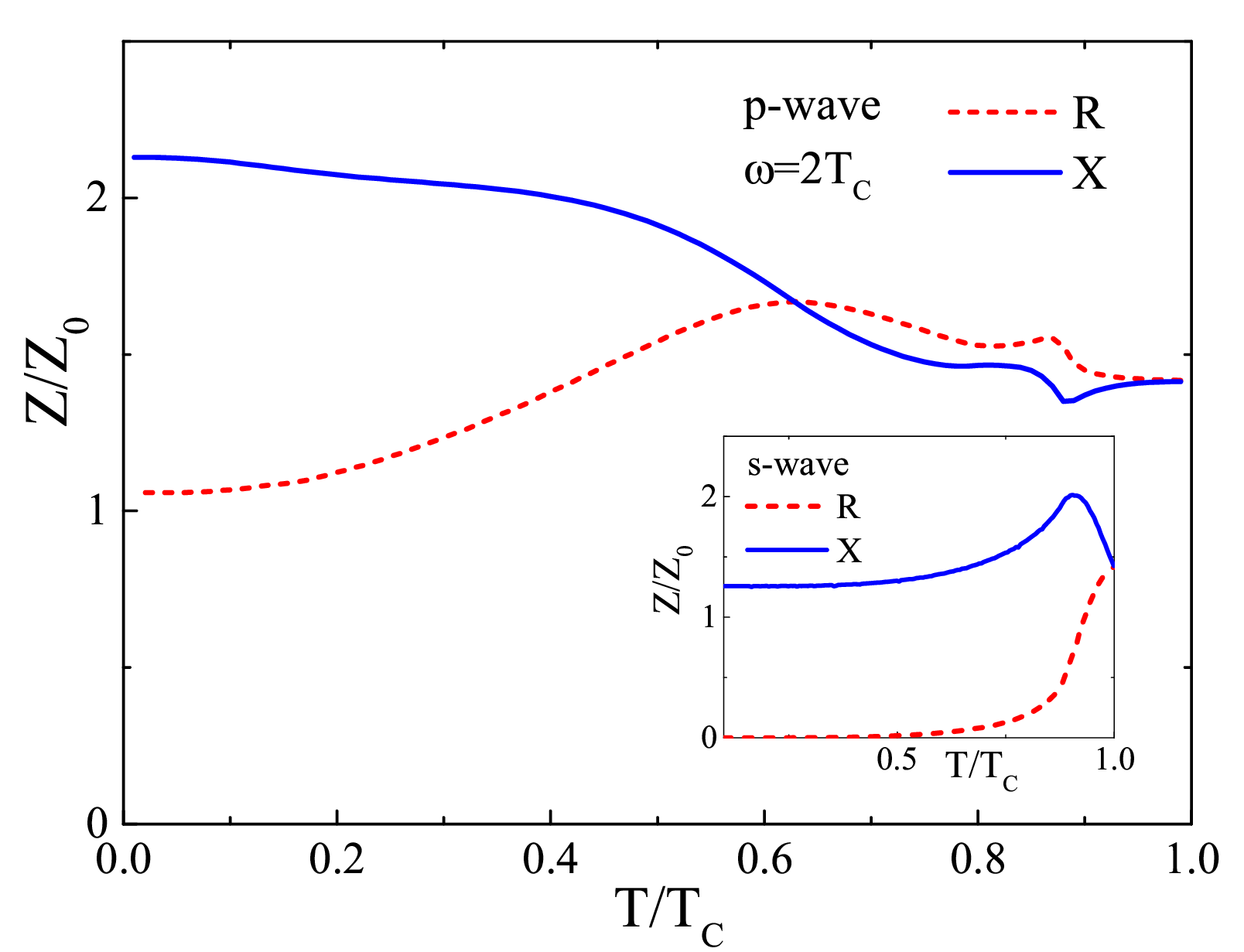}
\end{minipage}
\caption{Real $R$ and imaginary $X$ components of the local impedance $Z$ at the
rough surface of chiral $p_x+i p_y$ superconductor vs.\ temperature $T$ for different frequencies: (a)~$\omega =  0.05 T_c$,  (b)~$\omega =  0.5 T_c$, (c)~$\omega = 1.0 T_c$, and (d)~$\omega = 2.0 T_c$. Insets demonstrate temperature dependencies of $Z$ at the same $\omega$ for $s$-wave superconductor.}
\label{Z_T}
\end{figure*}

\subsection{Results for the surface impedance}

The local impedance $Z=R-iX$ is calculated from the local value of the complex conductivity $\sigma$ as
\begin{equation} \label{Imp}
Z(\omega)=\sqrt{\frac{4\pi\omega}{i c^2 \sigma(\omega)}},
\end{equation}
and we are interested in the surface value of this quantity (at the outer surface of the system). The local impedance, characterizing surface properties and effect of surface states, will later be used for comparison with experiment.

The standard theory of superconductivity prescribes a certain relation, $R<X$, between the real and imaginary parts of the surface impedance (i.e., between the surface resistance $R$ and the surface reactance $X$) \cite{Tinkham}.
This conventional relation is straightforwardly obtained if $\sigma_2 >0$. However, if $\sigma_2<0$, Eq.\ (\ref{Imp}) predicts an anomalous regime with $R>X$. This regime can be achieved locally due to odd-frequency superconductivity induced at the surface \cite{Asano-Fominov}.

\subsubsection{Frequency dependence of the surface impedance}

In order to trace how features of $Z(\omega)$ follow from features of $\sigma(\omega)$, we plot the impedance in Fig.~\ref{Z_hw} at the same parameters for which the conductivity was plotted in Fig.~\ref{Sigma}. The surface Andreev states, corresponding to $n_s<0$, lead to anomalous low-frequency behavior of the impedance, with $R>X$. This relation cannot originate from conventional bulk superconductivity with $n_s>0$ and is a manifestation of odd-frequency superconductivity \cite{Asano-Fominov}. At $\omega\sim \delta$, the anomalous Andreev-band contribution is already significantly reduced, $\sigma(\omega)$ reaches its maximum, and $Z(\omega)$ is in the conventional regime, with $R<X$. Then, at $\omega\approx 2\delta$, the intersection occurs, $R=X$, due to $\sigma_2(\omega)$ crossing zero. Finally, at even larger frequencies, $\sigma_2(\omega)$ remains small [$|\sigma_2(\omega)| \ll \sigma_1(\omega)$], and $Z(\omega)$ demonstrates the normal-metallic-like regime with
\begin{equation} \label{marganom1}
R(\omega) \approx X(\omega) \approx \sqrt{\frac{2\pi\omega}{c^2 \sigma_1(\omega)}}
\end{equation}
[which becomes truly normal metallic at $\omega$ exceeding $2\Delta_0$, in which case $\sigma_1(\omega) \approx \sigma_0$]. This regime is marginally anomalous, since $R$ slightly exceeds $X$ due to the negative value of $\sigma_2(\omega)$:
\begin{equation} \label{marganom2}
R(\omega) - X(\omega) \approx -\frac{\sigma_2(\omega)}{\sigma_1(\omega)} R(\omega).
\end{equation}

Comparison between Figs.~\ref{Z_hw}(a) and~\ref{Z_hw}(b) demonstrates that the low-frequency interval, in which the anomalous regime with $R>X$ takes place, shrinks when temperature is increased. This agrees with a general tendency to decreasing characteristic superconducting energy scales, which we previously noted in the behavior of conductivity.

The behavior of $Z(\omega)$ at $\omega\to 0$ can be extracted from the results for the complex conductivity $\sigma(\omega)$. According to Sec.~\ref{subsec:conductivity}, in this limit, the real part of conductivity turns to a (positive) constant $\sigma_1$, see Eq.\ (\ref{sigma1(0)}) (the characteristic scale of this constant is set by the Drude conductivity $\sigma_0$), while the imaginary part diverges as described by Eq.\ (\ref{sigma2}) with $n_s<0$.
Equation (\ref{Imp}) then leads to
\begin{equation} \label{RX}
R = \sqrt{\frac{\pi \tau}{\sigma_0} \frac n{|n_s|}} \frac{2\omega}{c},
\qquad
X = \sqrt{\frac{\pi \tau^3}{\sigma_0^3} \frac{n^3}{|n_s|^3}} \frac{\sigma_1 \omega^2}{c}.
\end{equation}

Therefore $R(\omega)$ is linear at $\omega\to 0$, while $X(\omega)$ is quadratic. This behavior can easily be seen in Fig.~\ref{Z_hw}(a), while the quadratic dependence of $X(\omega)$ is not so clear in Fig.~\ref{Z_hw}(b). The reason is that the low-frequency regime corresponding to Eq.\ (\ref{RX}) is rather narrow in Fig.~\ref{Z_hw}(b), since higher temperature weakens the effect of the Andreev band and leads to relatively small $n_s$ in this case, so that the coefficient in Eq.\ (\ref{RX}) is large.

Note that Eq.\ (\ref{RX}) confirms the anomalous $R>X$ relation at low frequencies (more precisely, this equation corresponds to $R\gg X$). This contrasts the conventional behavior in an $s$-wave superconductor with positive $n_s$. In this case, the low-frequency result can be written in the same form as Eq.\ (\ref{RX}) but with $R$ and $X$ interchanged (which results in the conventional relation $R<X$). Another quantitative difference is that $\sigma_1$ in the gapped $s$-wave superconductor is exponentially suppressed at low temperatures as $\exp(-\Delta_0/T)$, while in the case of Andreev band, $\sigma_1$ does not have this smallness.

\subsubsection{Temperature dependence of the surface impedance}

In experiments studying the surface impedance, frequency $\omega$ (or a set of frequencies) of the external electromagnetic field is usually fixed by a resonator and changing $\omega$ can be a challenging task. At the same time, temperature can be varied rather easily in a continuous manner. Therefore it is important to study $Z(T)$ at fixed $\omega$.

Figure~\ref{Z_T} demonstrates temperature dependences of real (dashed line) $R$ and imaginary
(solid line) $X$ part of the surface impedance $Z$ at different frequencies $\omega$ of the electromagnetic field. For comparison, we also plot (as insets in Fig.~\ref{Z_T} at the same $\omega$) the temperature dependences of the surface impedance for the case of rough surface of $s$-wave superconductor (in this case, we can simply assume the whole volume of the $s$-wave superconductor to be diffusive).

At lowest frequencies, Eq.\ (\ref{RX}) is applicable (this is the case of $\sigma_2<0$), where the temperature dependence enters through $n_s(T)$. For this anomalous regime to apply, transitions should be all inside the Andreev band, so $\omega$ should be at least much smaller than the band gap $\Delta_0-\delta$. This condition is met at almost all temperatures (except close vicinity of $T_c$) in the case of Fig.~\ref{Z_T}(a), which is plotted at $\omega = 0.05 T_c$. In accordance with Eq.\ (\ref{RX}), we obtain the anomalous regime with $R>X$. Both $R$ and $X$ grow with increasing temperature due to suppression of $n_s(T)$. In the conventional superconductor (see inset), $R<X$ in the whole superconducting temperature range.

As $\omega$ is increased, anomalous low-frequency contributions from the Andreev band to $\sigma_2$ are suppressed, while conventional ones (involving the bulk band) are enhanced, so that $\sigma_2$ can change its sign becoming positive. At $\omega = 0.5 T_c$ [see Fig.~\ref{Z_T}(b)], the balance of different contribution changes as a function of temperature, and we find a crossover between the anomalous and conventional regimes as $T$ grows [intersection of $R(T)$ and $X(T)$ at $T$ slightly lower than $0.6 T_c$]. This happens mainly because increasing temperature suppresses anomalous contributions from the Andreev band to $\sigma_2$, while $\omega$ is fixed [in addition, the superconducting energy scales $\Delta_0(T)$ and $\delta(T)$ decrease], so temperature influences relative magnitude of anomalous and conventional contributions to $Z(\omega)$.

At higher frequencies [see the case of $\omega = 1.0 T_c$ in Fig.~\ref{Z_T}(c)], the anomalous low-temperature regime, originating from the anomalous low-frequency behavior of $\sigma_2(\omega)$, disappears. We can say that the intersection point that we observed in Fig.~\ref{Z_T}(b) at $T\approx 0.6 T_c$, moves to the left and finally reaches zero temperature, after which the intersection disappears.
As a result, the reactive component of impedance $X$ exceeds $R$ in most parts of the superconducting temperature range, except for a narrow region near the critical temperature, $T>0.9 T_c$. In the latter case, the marginally anomalous normal-metallic-like regime [with $R$ only slightly exceeding $X$, as described by Eqs.\ (\ref{marganom1}) and (\ref{marganom2})] is due to the high-frequency marginally anomalous behavior of $\sigma_2(\omega)$, which we observed in Fig.~\ref{Sigma} at $\omega> 2\delta$. At fixed $\omega$, this regime is reached due to shrinking of $\delta$ with increasing $T$.

Finally, at even higher frequencies of the order of a few $T_c$ [see the case of $\omega = 2.0 T_c$ in Fig.~\ref{Z_T}(d)], the marginally anomalous normal-metallic-like regime fills wide temperature interval below $T_c$.
Small cusps around $T=0.9 T_c$ in Fig.~\ref{Z_T}(d) are manifestations of sharp peaks in the spectral structure of the Green functions at $|E|=\Delta_0$ (see Fig.~\ref{FG func}). The cusps correspond to the temperature, at which $\omega \approx 2 \Delta_0(T)$, so that transitions occur between two peaks.
At low temperatures, $R$ decreases while $X$ grows with decreasing $T$.
This behavior is significantly different from the $Z$ dependence in the $s$-wave superconductor, in which $X$ demonstrates the coherent peak below $T_c$ (due to transitions between the spectral peaks in the Green functions) and then decreases at low temperatures, as shown in the inset in Fig.~\ref{Z_T}(d).

In principle, the low-temperature behavior of $R(T)$ and $X(T)$ similar to Fig.~\ref{Z_T}(d) can be obtained in the conventional $s$-wave superconductor. However, frequencies $\omega$ required to achieve such behavior in the $s$-wave case, are markedly larger than the gap $2 \Delta_0$ in the quasiparticle spectrum. At the same time, presence of the Andreev band at the surface of the chiral $p$-wave superconductor leads to appearance of these effects at subgap frequencies $\omega < 2\Delta_0$.

As demonstrated by comparison between the main plots and insets in Fig.~\ref{Z_T}, the impedance behaves qualitatively and quantitatively differently in the cases of chiral $p$-wave and $s$-wave superconductors. The main special feature of the chiral $p$-wave case is the possibility of the anomalous regime with $R>X$. Another important feature is the strongly enhanced active part of the impedance: $R$ can be finite and even of the order of the normal-metallic value at subgap frequencies even at $T=0$, in contrast to zero $R$ in the $s$-wave case under the same conditions.
Both these features arise due to a significant amount of low-energy states at the surface of the chiral $p$-wave superconductor (the Andreev band).

\section{Experiment \label{Sec2}}

Three  high-quality single crystals of Sr$_{2}$RuO$_{4}$ were grown \cite{Kashiwaya} by the floating-zone method. The crystals were cut into small cylinders and then cleaved along the $ab$ plane. The surfaces were polished with diamond slurry to make flat thin pellets of Sr$_{2}$RuO$_{4}$. They are of the shape of approximately rectangular plates with the $ab$ planes on the major flat faces and thin edge surfaces. The exact dimensions are $1.2 \times 0.5 \times 0.04$,  $0.5 \times 0.3 \times 0.04$, and $0.7 \times 0.4 \times 0.03$~mm$^3$, respectively.

The temperature dependence of the surface impedance was measured by the ``hot finger'' technique in a cylindrical niobium cavity resonator. The walls of the resonator are cooled down with liquid helium and are in the superconducting state. The first experimental setup utilizes the resonator at frequency 9.4~GHz at the H$_{011}$ mode, and the second experimental setup works at 28.2~GHz at the H$_{011}$ mode and at 41.7~GHz at the H$_{013}$ mode \cite{Shev}.
The crystal was placed at the end of a sapphire rod in uniform high-frequency magnetic field.
The temperature of the rod and the sample can be varied in the range 5~K to 300~K in the first experimental setup and in the range 0.5~K to 100~K in the second setup.
Therefore, we investigated the superconducting and normal states of the samples in the 28.2~GHz and 41.7~GHz setup, while measurements in the 9.4~GHz setup were performed to probe the frequency dependence of impedance and to expand the temperature interval in the normal state.

\begin{figure}[t]
\centerline{\psfig{figure=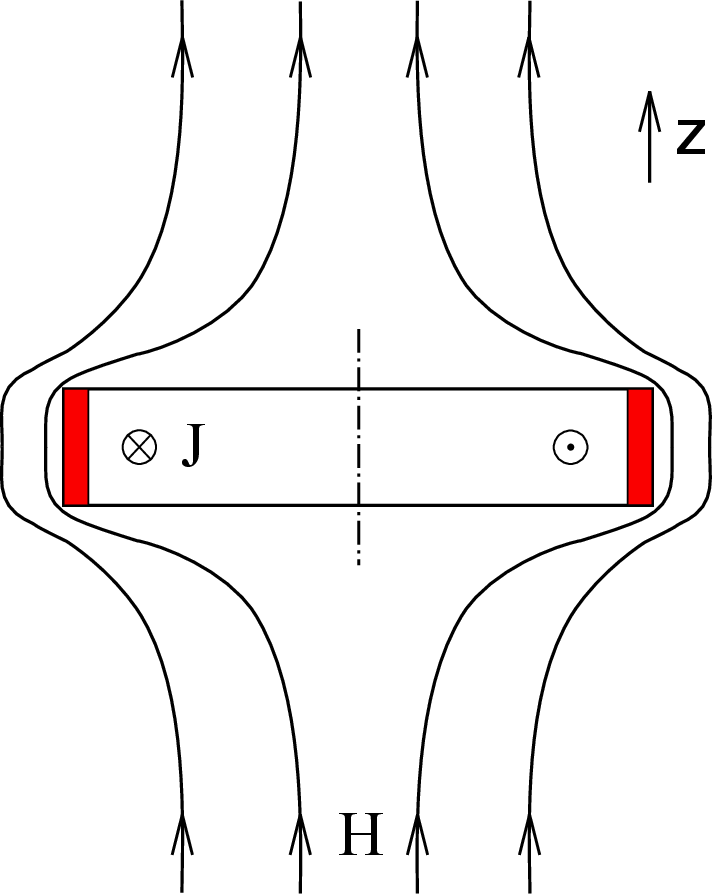,width=4cm,clip=,angle=0.}}
\caption{Schematic geometry of the sample. The major flat faces of the sample are perpendicular to the high-frequency magnetic field $H$ (which is directed along the $c$ axis).
The induced microwave currents $J$ flow parallel to the $ab$ planes. The magnetic field is largest near the edge surfaces of the sample (shown in red).
}
\label{Exp:SampleGeometry}
\end{figure}

The flat faces of the sample were perpendicular to the high-frequency magnetic field, see Fig.~\ref{Exp:SampleGeometry}.
In this orientation, the induced microwave currents flow parallel to the $ab$ planes on all six sides of a sample. In a thin plate, the microwave energy absorption is almost the same on the major flat faces and the edge surfaces of the sample.

The surface impedance of the sample, $Z(T)=R(T)-iX(T)$ is composed by the surface resistance $R(T)$ and the surface reactance $X(T)$. The surface impedance of the sample can be directly found from the measured experimental temperature dependences of the resonance frequency $f(T)$ and the quality factor $Q(T)$ of the resonance system. Using the perturbation theory, we find $R(T)=\Gamma/Q(T)$ and $\Delta X(T)=-2\Gamma f(T)/f(0)$, where $\Gamma$ is a coefficient of proportionality, which depends on the distribution of the electromagnetic field on the surface of the sample, and $\Delta X(T)$ is the \emph{change} in surface reactance as a function of temperature.
To find the \emph{absolute value} of $X$, additional information is required; for example, one may take into account that $R(T) = X(T)$ under the normal skin effect conditions. Usually, the slopes of the $R(T)$ and $\Delta X(T)$ curves coincide in a wide range of temperatures just above the superconducting transition point. Then, vertically shifting the $\Delta X(T)$ curve and superimposing it on $R(T)$, we can find $X(T)$.

\begin{figure}[t]
\centerline{\psfig{figure=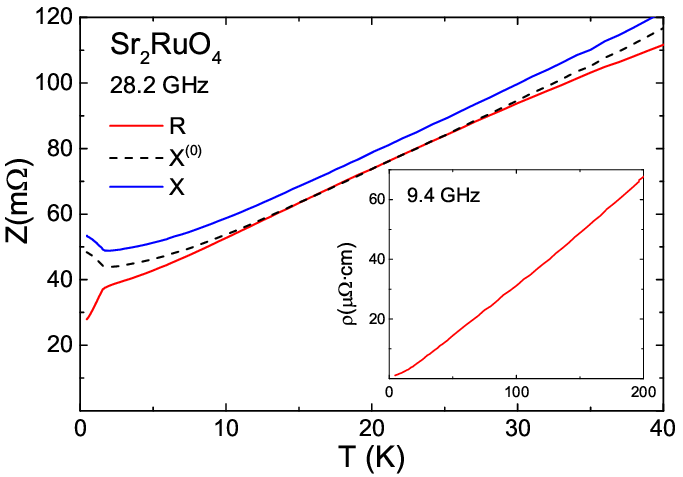,height=6.5cm,width=\columnwidth,clip=,angle=0.}}
\caption{Temperature dependence of the surface impedance in a Sr$_{2}$RuO$_{4}$ crystal at frequency 28.2~GHz.
The experimentally measured quantities are the surface resistance $R(T)$ and change in reactance $\Delta X(T)$.
Shifting $\Delta X(T)$ in order to achieve the best possible coincidence with $R(T)$ (solid red curve) in the normal state, we obtain $X^{(0)}(T)$ (dashed black curve). A more accurate result for the surface reactance, $X(T)$ (solid blue curve), is then found according to the procedure described in the main text. Inset: temperature dependence of the resistance, $\rho (T)= R^{2}(T) c^2/2\pi \omega$, extracted from measurements of $R(T)$ in the normal state of Sr$_{2}$RuO$_{4}$ at frequency 9.4~GHz.}
\label{Exp2}
\end{figure}

Experimentally, in our Sr$_{2}$RuO$_{4}$ crystals at 28.2~GHz, the slopes of the measured $R(T)$ and $\Delta X(T)$ curves coincide in the temperature range $15~\mathrm{K} <T<30$~K.
Then, we can shift our measured $\Delta X(T)$, obtaining the $X^{(0)}(T)$ curve that overlaps with $R(T)$ in this temperature range; this results in the dashed black curve in Fig.~\ref{Exp2} (the figure shows the results for one of the samples; the results on the two other samples were similar).
At $T>30$~K, the reactance curve $X^{(0)}(T)$ deviates from $R(T)$ (becomes larger).
This discrepancy can be explained by the temperature expansion of the sample in our experimental range up to 100~K \cite{ThermExp,ThermExp2}.

The inset in Fig.~\ref{Exp2} shows the temperature dependence of the resistivity calculated as $\rho (T)= R^{2}(T) c^2/2\pi \omega$ at 9.4~GHz setup. The resistivity obtained at frequency 28.2~GHz in the interval  $10~\mathrm{K}<T<100$~K is the same within experimental accuracy.

The resistivity can be described as $\rho (T)=(-3 + 0.35 \times T)$~($\mu\Omega$\,cm)  in the interval $20~\mathrm{K}<T<200$~K, and varies from 5~$\mu\Omega\,\mathrm{cm}$ at 20~K to 100~$\mu\Omega\,\mathrm{cm}$ at 300~K, in agreement with the results reported previously in Refs.\ \cite{resist1,resist2}.
From our measurements, the temperature dependence of the resistivity is almost linear. Note that this dependence cannot be explained by electron-phonon scattering according to the Bloch-Gr\"{u}neisen law, since this would imply unrealistically low Debye temperature $\sim 20$~K, which contradicts the measured value of $T_D \sim 460$~K \cite{TDebay}.

At low temperatures $T<20$~K, our results demonstrate some unexpected features. The measured value $\rho(4\,\mathrm{K})=1.5~\mu\Omega\,\mathrm{cm}$ is significantly higher than $\rho(4\,\mathrm{K})=0.25~\mu\Omega\,\mathrm{cm}$ reported in Ref.\ \cite{r0}.  Also, according to our high frequency measurements, $\rho \propto T^{1.5}$ below 20~K.

It should be emphasized that the electromagnetic technique (ac measurements) probes only a thin skin layer of the material, while material characteristics (chemical composition, etc.) in this layer may differ from the bulk. According to the measurements of the surface impedance in Refs.~\cite{Gough,Ormeno}, the resistance of different samples with the same superconducting transition temperature can differ by a factor of 1.5 at 10~K, and at $T<12$~K, the resistance has a power-law temperature dependence with exponent between 1.7 and 1.9.

Note also that our low-temperature results for $\rho(T)$ differ from the generally accepted quadratic electron-electron scattering law $\rho \propto T^{2}$ (based on dc measurements)  \cite{resist1}.

Surprisingly, below 15~K, the reactance $X^{(0)}(T)$ becomes larger than the surface resistance $R(T)$.
This feature cannot be explained within the framework of the normal skin effect, and this subtlety necessitates modifications of the usual procedure for calculating the reactance $X(T)$ from the measured $R(T)$ and $\Delta X(T)$ dependences.
In Refs.\ \cite{Gough,Ormeno}, at frequencies below 15~GHz, a similar behavior of the surface impedance was observed, and it was attributed to not too small values of the relaxation time $\tau_0$.
While typically $\omega\tau_0 \ll 1$ in our experimental range of frequencies,
a natural reason for a difference between $X$ and $R$ can still be a small but finite value of $\omega\tau_0$
(inside the skin layer).
In this case, even in the normal state, the current lags behind the electric field, and $X>R$ even in the normal state.
Indeed, in the local limit, according to the Drude model, the microwave conductivity is $\sigma_{\omega} = \sigma_0/(1-i\omega\tau_0)$. In this model, the impedance components are related to each other as $X_\mathrm{fit}=\sqrt{R^2+(Z_v \omega/\omega_p)^2}$, where $Z_v=120\pi~\Omega$ is the vacuum impedance and $\omega_p=\sqrt{4\pi n e^2/ m}$ is the plasma frequency of the material. The plasma frequency can be considered as a fitting parameter determining the overall slope of $X_\mathrm{fit}(T)$.
Having adjusted the overall slope of $X_\mathrm{fit}(T)$ to that of $\Delta X(T)$, we can use our freedom of shifting $\Delta X(T)$ in order to achieve agreement with $X_\mathrm{fit}(T)$ in a widest possible range of temperatures.
This results in $X(T)$ shown as the blue curve in Fig.~\ref{Exp2} [the $X_\mathrm{fit}(T)$ curve is not shown]. The $X(T)$ curve is thus a more accurate result for the reactance, compared to $X^{(0)}(T)$ discussed before.

The value of $\omega_p=4.5~\mathrm{eV}$, calculated from the band structure \cite{Oguchi}, does not fit our $Z(T)$ below 15~K. The fit is much better if we assume $\omega_p=1.5~\mathrm{eV}$, as proposed in Ref.\ \cite{Zabolotnyy};
then, some discrepancy within $10\%$ between $X(T)$ and $X_\mathrm{fit}(T)$ is observed only below 7~K.
In the local limit, we have $\omega \tau_0=(X^2-R^2)/2XR$ and
measurements at frequency 28.2~GHz near the superconducting transition temperature
give $\omega \tau_0 \simeq 0.25$.

Another reason for discrepancy between $X$ and $R$ in the normal state can be related to deviations from the local-limit conditions in the skin layer (this limit implies $l_0 \ll \delta_0$, with $l_0$ and $\delta_0$ being the mean free path and the skin depth, respectively). For example, in the extreme nonlocal limit, when
$l_0 \gg \delta_0$, the anomalous skin effect regime with $X=\sqrt{3}R$ is realized.
The mean free path $l_0$ can be estimated from the residual resistivity \cite{resist1} and for $\rho(4\,\mathrm{K})=1.5~\mu\Omega~\mathrm{cm}$, we find $l_0 \approx 200$~nm
and $\delta_0 \approx 350$~nm. So, our measurements below 7~K are performed not exactly in the local but rather in the intermediate regime with $l_0 \lesssim \delta_0$, and deviations in the fitting between $X(T)$ and $X_\mathrm{fit}(T)$  can be associated with nonlocality \cite{nonlocal}.

\begin{figure}[t]
\centerline{\psfig{figure=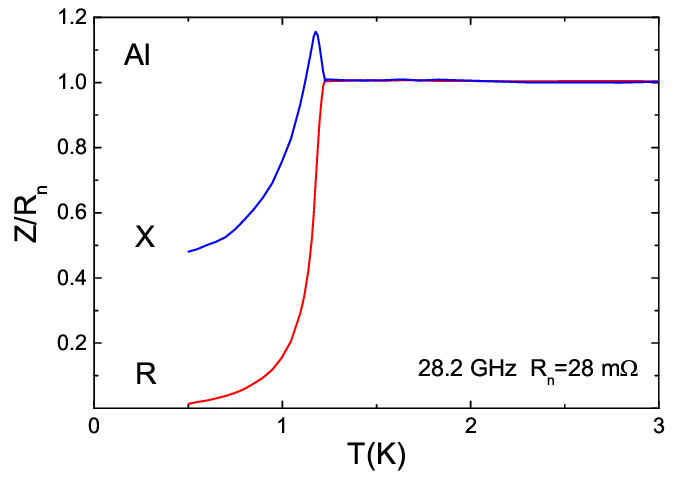,height=6.5cm,width=\columnwidth,clip=,angle=0.}}
\caption{Conventional temperature dependence of the surface impedance $Z = R - iX$: aluminum crystal at frequency 28.2~GHz.}
\label{Exp1}
\end{figure}

\begin{figure}[t]
\centerline{\psfig{figure=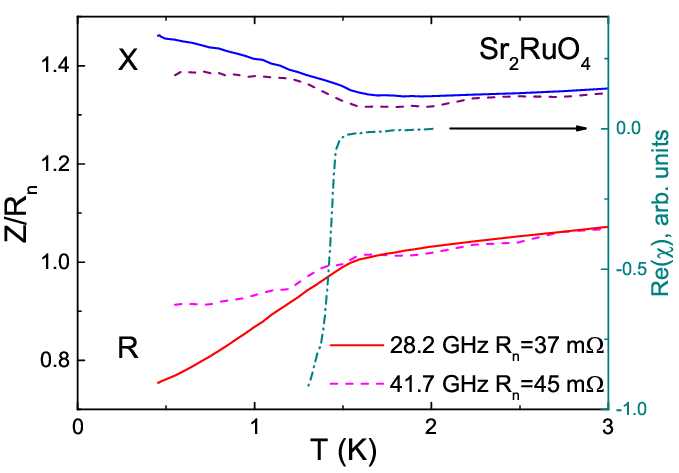,height=6.5cm,width=\columnwidth,clip=,angle=0.}}
\caption{Left axis: temperature dependences of the surface impedance of Sr$_{2}$RuO$_{4}$ crystal, normalized to $R_n=R(1.5\,\mathrm{K})$, at low temperatures at frequencies 28.2~GHz (solid curves) and 41.7~GHz (dashed curves).  Right axis:  temperature dependence of the dynamic magnetic susceptibility at frequency 100~kHz (dash-dotted curve).
Taking $T_c = 1.5$~K, we find that our frequencies 28.2~GHz and 41.7~GHz correspond to $\omega \approx 0.9 T_c$ and $\omega \approx 1.3 T_c$, respectively.}
\label{Exp3}
\end{figure}

Before studying the samples of Sr$_{2}$RuO$_{4}$ in the superconducting state, let us consider a typical picture for the temperature dependence of the surface impedance in the case of a conventional superconductor.
Figure~\ref{Exp1} shows results of measurements on an aluminum sample with dimensions $0.5 \times 0.4 \times 0.13$~mm$^3$.  The following features can be noted: immediately above the superconducting transition temperature, real $R(T)$ and imaginary $X(T)$  parts of the surface impedance coincide; below $T_c=1.2$~K,
the real part of impedance $R(T)$ decreases almost to zero already at $T=T_c/2$, while the imaginary part of impedance $X(T)$ has a small hump just below $T_c$ (associated with the coherence peaks in the spectral characteristics of the superconductor), and then $X(T)$ decreases monotonically.

To verify that the samples of Sr$_{2}$RuO$_{4}$ are indeed in the superconducting state, we measured the temperature dependence of the dynamic magnetic susceptibility $\chi (T)$ at a frequency of 100~kHz. The $\re \chi (T)$ curve in Fig.~\ref{Exp3} (relating to the right axis) clearly indicates a sharp shielding of the electromagnetic field below 1.5~K and hence the superconducting transition with $T_c = 1.5$~K.

We were able to measure the impedance in the superconducting state only at 28.2~GHz and 41.7~GHz frequencies due to temperature limitations of our experimental setup.
%Additional measurements were made at low temperatures in the same setup at the H$_{013}$ mode  at frequency 41.7~GHz.
At both frequencies, a peculiar feature of our measurements is that with temperature decreasing below 1.5~K, the real part of the impedance $R(T)$
decreases very smoothly extrapolating to a large zero-temperature value
(contrary to what is expected for a conventional $s$-wave superconductor at subgap frequencies, $\omega<2\Delta_0$), see Fig.~\ref{Exp3}. At the lowest temperature of 0.5~K, the $R$ value is $75\%$ of $R(T_c)$ at 28.2~GHz and $90\%$ of $R(T_c)$ at 41.7~GHz.
The reactance $X(T)$
monotonically grows below 1.5~K down to the lowest temperature of 0.5~K at both frequencies.

\section{Discussion}
\label{Sec3}

The main qualitative feature of our experimental results for the surface impedance of Sr$_{2}$RuO$_{4}$ samples is large values of the surface resistance $R$ down to low temperatures at subgap frequencies, $\omega < 2\Delta_0$ (in out experimentally conditions, the subgap frequencies are those below approximately 110~GHz). This is clear evidence of a significant amount of subgap states in superconducting Sr$_{2}$RuO$_{4}$.

The theoretical model developed in this work actually explains the appearance of such states, if the bulk pair potential has the chiral $p$-wave symmetry. In this case, we predict that odd-frequency $s$-wave triplet superconductivity is generated at the surface providing a broad band of subgap Andreev bound states (see Fig.~\ref{FG func}).
An important ingredient of the model is the assumption about surface roughness, which leads to isotropisation of superconductivity in the vicinity of the surface, thus enhancing the odd-frequency $s$-wave component that is manifested in formation of the surface subgap states \cite{Nagato2011,Bakurskiy1}.
We expect such an assumption to be realistic for the studied Sr$_{2}$RuO$_{4}$ samples.

Figure~\ref{Exp3} that summarizes experimental data at two different frequencies, shows a very peculiar trend, namely, quite large values of surface resistance $R$ which increase with frequency.
This behavior is clearly inconsistent with the standard $s$-wave superconductivity (see Fig.~\ref{Exp1}) but is consistent with the results of calculations for the $p$-wave case
shown in Figs.~\ref{Z_T}(c) and~\ref{Z_T}(d). Similar experimental results were reported earlier in Refs.\ \cite{Gough,Ormeno,Baker} and have been interpreted in the framework of a phenomenological two-fluid model assuming a significant ``normal-fluid'' fraction at low $T$. Therefore our model provides a microscopic background for such a two-fluid model.

The surface reactance $X(T)$, monotonically growing with decreasing temperature, also does not agree with the expected behavior in the $s$-wave case. In the latter case, we would expect a coherent peak just below $T_c$ followed by a monotonic decay (see experimental Fig.~\ref{Exp1} and theoretical curves in the insets in Fig.~\ref{Z_T}).

Nevertheless, note that the surface impedance behavior, qualitatively similar to Figs.~\ref{Exp3} and~\ref{Z_T}(d), is possible in the $s$-wave case, but only at frequencies essentially exceeding the superconducting gap. This is however clearly not the situation that we have in our experiment.

Qualitatively, the experimental results for the surface impedance, Fig.~\ref{Exp3}, are best described by the theoretical curves shown in Fig.~\ref{Z_T}(d). At the same time, the parameters of Fig.~\ref{Z_T}(d) correspond to frequency $\approx 60$~GHz, which is larger than experimental frequencies in Fig.~\ref{Exp3}. In this respect, we should note that our theoretical results cannot provide real quantitative fitting of experiment due to several reasons. First, we calculate only the surface contribution to the impedance, originating from the rough surface layer. In our model, this layer is assumed to be very thin. If the skin depth is larger than the rough layer's thickness, there is also a bulk contribution from clean anisotropic superconductivity. Second, the geometry of the experimental setup is such that the edge surfaces of the sample provide only about one half of the whole microwave absorption, while the other half is due to major flat faces of the sample. Theoretically, we consider a single surface with Andreev bound states; this corresponds to the edge surfaces of experimental samples. We therefore do not capture the contribution to the surface impedance originating from major flat faces in experiment. Finally, we do not consider possible surface current in the chiral $p$-wave state \cite{Suzuki,Bakurskiy1}.

The above theoretical simplifications were made in order to focus on and underline the effect of the surface Andreev subgap states and the corresponding anomalies in the surface impedance. We find that theoretically calculated anomalies are in qualitative agreement with experiment. At the same time, full theoretical description of the performed experiments requires advancing theory further in order to take into account the above-mentioned effects that are missing in our consideration.

While experiment qualitatively agrees with features of the surface impedance described by our theory, it does not show indications of the most interesting anomalous regime, in which the relation between the surface resistance and impedance is inverted and becomes $R>X$. Theoretically, we find this regime at low temperatures and frequencies [see Figs.~\ref{Z_T}(a) and~\ref{Z_T}(b)], when the anomalous effect from the odd-frequency superconductivity and the surface Andreev band is maximized. In order to maximize this contribution, the conditions $T\ll \delta$ and $\omega \ll \Delta_0-\delta$ should be met (then the Andreev band fully participates in microscopic transitions defining the microwave response, and all the transitions are within the Andreev band).

Finally, we comment upon applicability of our theoretical results to different triplet states. Above, we have focused on the chiral $p$-wave superconductivity that breaks the time-reversal symmetry. This is the main candidate for the superconducting state in Sr$_{2}$RuO$_{4}$ since there are experimental indications that the time-reversal symmetry is indeed broken in this material \cite{Kittaka}. At the same time, the time-reversal-invariant helical $p$-wave superconducting state is not completely excluded, so this alternative is also under discussion \cite{Kittaka,Tada2009,Scaffidi2015,Kawai2017}.
In this respect, we note that our results can be applied to the helical $p$-wave case as well, since
apart from the features related to edge currents, the structure of the subgap Andreev states is the same in the two cases
(in the absence of spin-orbit coupling and exchange fields).

\section{Conclusions}
\label{sec:conclusions}

In this work, we have calculated local impedance at a diffusive surface of a chiral $p$-wave superconductor. Our theoretical approach is based on the quasiclassical Eilenberger-Larkin-Ovchinnikov formalism, where the spatial dependence of the pair potential is determined self-consistently. The obtained real and imaginary components of the Green functions (pairing functions and the local density of states) demonstrate an energy dependence featuring the subgap band of dispersive Andreev states.
Using the obtained solutions for the Green functions, we have calculated the local complex conductivity and surface impedance of the chiral $p$-wave superconductor in a broad range of microwave frequencies (from subgap to above-gap regime).

We have identified anomalous features of the complex conductivity and surface impedance caused by generation of odd-frequency pairing at the surface. The low-frequency microwave response turns out to be anomalous due to contributions from the Andreev band (which is a manifestation of the odd-frequency pairing). The odd-frequency symmetry locally leads to the change of sign of $n_s$, the coefficient determining the supercurrent response to external field and usually interpreted as the density of superconducting electrons. The Andreev-band contribution at low frequencies makes $n_s$ negative near the surface.

As a result, the imaginary part of the local conductivity also becomes negative, which leads to anomalous relation, $R>X$, between the local surface resistance and reactance. This effect takes place at frequencies several times smaller than the bulk superconducting gap; the anomalous range of frequencies is sensitive to $\Delta_0$ and $\delta$, the half-widths of the bulk gap and Andreev band, respectively. The anomalous range of frequencies is maximized at low temperatures, $T\ll T_c$.

The obtained theoretical results are compared with experiments on surface impedance of Sr$_{2}$RuO$_{4}$ at frequencies 28.2~GHz and 41.7~GHz.
Experimentally, the temperature dependences of $R(T)$ and $X(T)$ show clear qualitative differences as compared to Al where conventional $s$-wave pairing is realized. Although the $R>X$ regime was not reached in experiment, the results are unconventional.
Our theoretical calculations demonstrate that experimentally observed anomalies of the surface impedance in Sr$_{2}$RuO$_{4}$ are consistent with the chiral $p$-wave scenario. One of the main experimental findings is strongly enhanced low-temperature surface resistance at subgap frequencies, which is incompatible with the conventional $s$-wave scenario. Our theoretical results provide microscopic explanation of enhanced subgap resistance in terms of the subgap Andreev states. In particular, this gives a microscopic basis to a phenomenological two-fluid model with enhanced normal component (finite quasiparticle fraction at $T=0$), previously employed to explain the impedance results in Sr$_{2}$RuO$_{4}$.

\acknowledgments

This work was supported by
the Ministry of Education and Science of the Russian Federation (Grant No.\ 14.Y26.31.0007) and
Japan-Russia research collaborative program (JSPS Bilateral Joint Research Projects and RFBR Grant No.\ 17-52-50080).
Ya.V.F.\ was also supported in part by the Basic research program of HSE.
The work of the Japanese participants was supported by
a Grant-in-Aid for Scientific Research on Innovative Areas, Topological Material Science
(Grant Nos.\ JP15H05851, JP15H05852, JP15H05853, and JP15K21717) and
a Grant-in-Aid for Scientific Research B (Grant No.\ JP15H03686 and No.\ JP18H01176)
from the Ministry of Education, Culture, Sports, Science, and Technology, Japan (MEXT).

\appendix

\section{Complex conductivity: derivation}
\label{app:derivation}

Here, we derive the expression for the local complex conductivity, Eqs.\ (\ref{sigma1}) and (\ref{sigma2}), in terms of the quasiclassical Usadel Green functions. Our derivation generalizes the one presented in Ref.~\cite{Glazman}, from the singlet case to the case of arbitrary spin structure of the superconducting state. Applying this approach to chiral $p$-wave superconductors, we are interested in the triplet superconducting state.

The nontrivial point in the latter case is that $p$-wave superconductivity cannot survive in the diffusive region, while the proximity-induced $s$-wave triplet superconducting correlations may still exist (odd-frequency superconducting state \cite{Tanaka2007a,Tanaka2007b,TanakaRev}).
Since in our system the bulk source of superconductivity is a clean $p$-wave superconductor, we start from the quasiclassical Eilenberger-Larkin-Ovchinnikov equations, which make it possible to trace the transition from clean to dirty limit (note that the diffusive normal layer modeling the rough surface can also be considered as a diffusive region of the same $p$-wave superconductor; in the limit of $\Delta \tau \ll 1$, the two models are equivalent).

\subsection{General pair potential}

The Eilenberger-Larkin-Ovchinnikov equation \cite{Eilenberger,LO1968} for the retarded Green function (in the real-energy representation) is formulated as
\begin{align}
& i v\hat{\mathbf{k}}
{\nabla}_{\mathbf{r}} \, \check{\mathrm{g}}^R
+\bigl[ \check{\mathcal{H}}^R, \check{\mathrm{g}}^R \bigr]=0,
\label{eilenberger_r2}
\\
& \check{\mathcal{H}}^R
=
\begin{pmatrix}
(E+i 0) \hat{\sigma}_0 &
\hat{\Delta}(\mathbf{r},\hat{\mathbf{k}},E) \\
 \undertilde{\hat{\Delta}}(\mathbf{r},\hat{\mathbf{k}},E) &
-(E+i 0) \hat{\sigma}_0
\end{pmatrix}
+
\frac{i}{2\tau} \bigl\langle \check{\mathrm{g}}^R(\mathbf{r},\hat{\mathbf{k}},E) \bigr\rangle_{\hat{\mathbf{k}}},
\\
& \check{\mathrm{g}}^R(\mathbf{r},\hat{\mathbf{k}},E)
=\begin{pmatrix}
\hat{\mathrm{g}}^R(\mathbf{r},\hat{\mathbf{k}},E) & \hat{\mathrm{f}}^R(\mathbf{r},\hat{\mathbf{k}},E) \\
-\undertilde{\hat{\mathrm{f}}}^R(\mathbf{r},\hat{\mathbf{k}},E) &
-\undertilde{\hat{\mathrm{g}}}^R(\mathbf{r},\hat{\mathbf{k}},E)
\end{pmatrix}.
\label{g_eilenberger_r2}
\end{align}
The Pauli matrices in spin space are denoted by $\hat{\sigma}_\nu$ with $\nu=0,1,2,3$, where $\hat{\sigma}_0$ is the unit matrix. The unit vector in the direction of the wave vector is represented by $\hat{\mathbf{k}}$. In all other instances, the hat accent ($\hat{\phantom{a}}$) denotes $2\times 2$ matrices, while the check accent ($\check{\phantom{a}}$) denotes $4\times 4$ matrices (in the direct product of spin and Nambu-Gor'kov spaces).
Throughout this section, we define the ``undertilde'' operation as
\begin{equation} \label{undertilde}
\undertilde{X}(\mathbf{r},\hat{\mathbf{k}},E) = X^* (\mathbf{r},-\hat{\mathbf{k}},-E).
\end{equation}
The Green function obeys normalization condition
$\left( \check{\mathrm{g}}^R \right)^2=\check{1}$.

The pair potential can generally be expanded into the spin components as
\begin{align} \label{DeltaAllComponents}
\hat{\Delta}(\mathbf{r},\hat{\mathbf{k}},E)
=\sum_{\nu} i \Delta_\nu(\mathbf{r},\hat{\mathbf{k}},E) \hat{\sigma}_\nu \hat{\sigma}_2,
\end{align}
with one singlet component ($\nu=0$) and three triplet components ($\nu=1,2,3$).
The $\Delta_\nu$ components are generally complex quantities.
We only consider the pair potentials belonging to the even-frequency symmetry class (the problematic of the odd-frequency pair potentials was recently discussed in Ref.~\cite{Fominov2015}),
\begin{align}
\Delta_\nu(\mathbf{r},\hat{\mathbf{k}},-E)=\Delta_\nu(\mathbf{r},\hat{\mathbf{k}},E).
\end{align}
In the absence of spin-dependent potentials and spin-orbit interaction, the normal part of the Green function is trivial in the spin space, while the anomalous part has the same spin components as the pair potential:
\begin{align}
\hat{\mathrm{g}}^R(\mathbf{r},\hat{\mathbf{k}},E)
&=
g^R(\mathbf{r},\hat{\mathbf{k}},E) \hat\sigma_0,
\\
\hat{\mathrm{f}}^R(\mathbf{r},\hat{\mathbf{k}},E)
&=
\sum_{\nu} i_s f^R_\nu(\mathbf{r},\hat{\mathbf{k}},E) \hat{\sigma}_\nu \hat{\sigma}_2,
\label{Fcomponents}
\end{align}
where
\begin{equation}
i_s =
\left\{
\begin{array}{cll}
1, & \text{spin-singlet} \; (\nu=0) \\
i, & \text{spin-triplet}\; (\nu=1,2,3).
\end{array}
\right.
\end{equation}
We intentionally introduce the spin-dependent $i_s$ factor and thus discriminate different spin components in the expansion (\ref{Fcomponents}). Our motivation is that this convention finally leads to the expression for the ac conductivity, Eq.\ (\ref{sigma_all}), that has exactly the same form in both the singlet and triplet cases [see also Eqs.\ (\ref{normal}) and (\ref{Us_all}), which have the same form in the two cases].

In the diffusive limit, the Eilenberger-Larkin-Ovchinnikov equation reduces to the Usadel equation \cite{Usadel1977} for $\check{\mathcal{G}}^R (\mathbf{r},E)$, the isotropic part of the Green function:
\begin{align}
& D \nabla_{\mathbf{r}} \left(
\check{\mathcal{G}}^R(\mathbf{r},E) {\nabla}_{\mathbf{r}}
\check{\mathcal{G}}^R(\mathbf{r},E)
\right)+
i \bigl[ \check{\mathcal{H}}_0^R ,\; \check{\mathcal{G}}^R(\mathbf{r},E)\bigr]=0,
\label{usadel_app2}
\\
&\check{\mathcal{H}}_0^R
=
\begin{pmatrix}
(E+i 0) \hat{\sigma}_0 &
i \Delta_0 (\mathbf{r},E) \hat\sigma_2 \\
i \Delta_0^* (\mathbf{r},E) \hat\sigma_2 &
-(E+i 0) \hat{\sigma}_0
\end{pmatrix},
\label{HR}
\\
&\check{\mathcal{G}}^R (\mathbf{r},E) = \bigl\langle \check{\mathrm{g}}^R (\mathbf{r},\hat{\mathbf{k}},E) \bigr\rangle_{\hat{\mathbf{k}}},
\label{UsadelGF}
\end{align}
where $D$ is the diffusion constant.
Note that in the diffusive limit, only the $s$-wave (hence, singlet) component of the pair potential can survive in Eqs.\ (\ref{usadel_app2})-(\ref{UsadelGF}). At the same time, the symmetry of the Green function can also correspond to the $s$-wave triplet odd-frequency superconducting correlations. In this case (realized, e.g., in the diffusive region of a $p$-wave superconductor \cite{Tanaka2007a,Tanaka2007b,TanakaRev}), the singlet pair potential $\Delta_0$ turns to zero in Eqs.\ (\ref{usadel_app2})-(\ref{UsadelGF}), while superconducting correlations are proximity-induced from a clean region of a triplet superconductor.

Similarly to Ref.\ \cite{Glazman}, one can generalize the Usadel equations (\ref{usadel_app2})-(\ref{UsadelGF}) to include a time dependent vector potential $\mathbf{A}(\mathbf{r},t)$ and consider the linear response to such a perturbation. The (complex) conductivity is then expressed in terms of the unperturbed Green functions as \cite{Glazman}
\begin{multline} \label{sigma}
\frac{\sigma(\mathbf{r},\omega)}{\sigma_0}
=-\frac{\pi}{8\omega} \int\frac{dE
}{2\pi}
\Tr
\bigl\{
\check{\mathcal{T}}_3\; \check{\mathcal{G}}^R(\mathbf{r},E+\omega)\;  \check{\mathcal{T}}_3\; \check{\mathcal{G}}^K(\mathbf{r},E)
\\
+\check{\mathcal{T}}_3\; \check{\mathcal{G}}^K(\mathbf{r},E+\omega)\;
  \check{\mathcal{T}}_3\; \check{\mathcal{G}}^A(\mathbf{r},E)
\bigr\},
\end{multline}
where
\begin{equation}
\check{\mathcal{T}}_3
=
\begin{pmatrix}
\hat\sigma_0 & 0 \\
0 & -\hat\sigma_0
\end{pmatrix}.
\end{equation}
The advanced ($\check{\mathcal{G}}^A$) and Keldysh ($\check{\mathcal{G}}^K$) Green functions can be written in terms of the retarded one:
\begin{align}
\check{\mathcal{G}}^A(\mathbf{r},E)
&=
-\check{\mathcal{T}}_3 \left( \check{\mathcal{G}}^R(\mathbf{r},E) \right)^\dagger \check{\mathcal{T}}_3,
\\
\check{\mathcal{G}}^K(\mathbf{r},E)
&=
\left(
\check{\mathcal{G}}^R(\mathbf{r},E) - \check{\mathcal{G}}^A(\mathbf{r},E) \right)
\tanh\Bigl(\frac{E}{2T}\Bigr).
\end{align}
Below, we will only deal with the retarded Green function and omit the corresponding $R$ superscript. Equation (\ref{sigma}) can be rewritten as
\begin{widetext}
\begin{align}
\frac{\sigma(\mathbf{r},\omega)}{\sigma_0}
=\frac{1}{16 \omega} \int dE
\Tr
\Bigl\{
&
\left(
\check{\mathcal{G}}^\dagger(\mathbf{r},E) \check{\mathcal{G}}(\mathbf{r},E+\omega) +
\check{\mathcal{G}}^\dagger(\mathbf{r},E) \check{\mathcal{T}}_3 \check{\mathcal{G}}^\dagger(\mathbf{r},E+\omega) \check{\mathcal{T}}_3
\right)
\tanh\Bigl(\frac{E+\omega}{2T}\Bigr)
\notag \\
&-
\left(
\check{\mathcal{G}}^\dagger(\mathbf{r},E) \check{\mathcal{G}}(\mathbf{r},E+\omega) +
\check{\mathcal{G}}(\mathbf{r},E) \check{\mathcal{T}}_3 \check{\mathcal{G}}(\mathbf{r},E+\omega) \check{\mathcal{T}}_3
\right)
\tanh\Bigl(\frac{E}{2T}\Bigr)
\Bigr\}.
\label{sigma_omega}
\end{align}
\end{widetext}

\subsection{Pair potential with a single spin component}

Up to now, we have worked with $4\times 4$ Green functions, assuming arbitrary (mixed) spin structure of the pair potential (\ref{DeltaAllComponents}). The above equations can be simplified to $2\times 2$ form if we only consider a single-component pair potential, i.e., if we assume that $\hat\Delta$ has only one spin component $\nu$:
\begin{equation} \label{DeltaSingleComp}
\hat{\Delta}(\mathbf{r},\hat{\mathbf{k}},E)
=i \Delta(\mathbf{r},\hat{\mathbf{k}},E) \hat{\sigma}_\nu \hat{\sigma}_2,
\end{equation}
where we omit the subscript $\nu$ in the scalar pair potential (in the right-hand side).
The definition of the undertilde operation, Eq.\ (\ref{undertilde}), immediately leads to the following relations:
\begin{multline}
\undertilde{\hat{\Delta}}(\mathbf{r},\hat{\mathbf{k}},E)
=
i \Delta^* (\mathbf{r},-\hat{\mathbf{k}},-E) \hat\sigma_\nu^* \hat\sigma_2
\\
= -i s_s s_p s_f \Delta^* (\mathbf{r},\hat{\mathbf{k}},E) \hat\sigma_2 \hat\sigma_\nu
= i \Delta^* (\mathbf{r},\hat{\mathbf{k}},E) \hat\sigma_2 \hat\sigma_\nu,
\end{multline}
where $s_s$, $s_p$, and $s_f$ are $\pm 1$ depending on the spin, parity, and frequency (energy) symmetry of $\Delta$, respectively. In the last equality, we have taken into account that the Pauli principle (fermionic antisymmetry of the Cooper pairing) requires $s_s s_p s_f = -1$.

In the case of a single-component pair potential, the anomalous part of the Green function also has a single component,
\begin{equation} \label{gSingleComp}
\check{\mathrm{g}}(\mathbf{r},\hat{\mathbf{k}},E)
=\begin{pmatrix}
g \hat\sigma_0 & i_s f \hat\sigma_\nu \hat\sigma_2 \\
i_s \undertilde{f} \hat\sigma_2 \hat\sigma_\nu &
-\undertilde{g} \hat\sigma_0
\end{pmatrix}_{(\mathbf{r},\hat{\mathbf{k}},E)},
\end{equation}
and we can ``disentangle'' the spin structure of this $4\times 4$ Green function by the following unitary transformation:
\begin{align}
\check{g} &=
\check{\Gamma}_\nu\; \check{\mathrm{g}} \; \check{\Gamma}_\nu^{-1},
\label{GammaRotation}
\\
\check{\Gamma}_\nu &=
\begin{pmatrix}
\hat{\sigma}_0 & 0 \\ 0 & \hat{\sigma}_\nu \hat{\sigma}_2
\end{pmatrix}.
\end{align}
The resulting Green function $\check{g}$ becomes spinless (proportional to $\hat\sigma_0$). Theory can then be reduced to a $2\times 2$ form in the Nambu-Gor'kov space, and the (retarded) $2\times 2$ matrix $\hat{g}$ obeys the following equations:
\begin{align}
&i v \hat{\mathbf{k}}
{\nabla}_{\mathbf{r}} \, \hat{g}
+\bigl[ \hat{H}, \hat{g} \bigr]=0,
\label{eilenberger_22}
\\
&\hat{H}(\mathbf{r},\hat{\mathbf{k}},E) =
\begin{pmatrix}
E+i0 &  i \Delta (\mathbf{r},\hat{\mathbf{k}},E) \\
i \Delta^* (\mathbf{r},\hat{\mathbf{k}},E) & -E-i0
\end{pmatrix}
+
\frac{i}{2\tau} \bigl\langle \hat{g}(\mathbf{r},\hat{\mathbf{k}},E) \bigl\rangle_{\hat{\mathbf{k}}},
\\
&\hat{g}(\mathbf{r},\hat{\mathbf{k}},E)=
\begin{pmatrix}
g & i_s f \\
i_s\undertilde{f} & - \undertilde{g}
\end{pmatrix}_{(\mathbf{r},\hat{\mathbf{k}},E)},
\end{align}
which are a reduced form of Eqs.\ (\ref{eilenberger_r2})-(\ref{g_eilenberger_r2}).
From the normalization condition, we obtain
\begin{align} \label{GundertildeG}
g^2 + i_s^2 f \undertilde{f} = 1,\quad g = \undertilde{g}.
\end{align}
In the Matsubara representation ($E = i\omega_n$), Eqs.\ (\ref{eilenberger_22})-(\ref{GundertildeG}) for the triplet case ($i_s=i$) lead to the Eilenberger-Larkin-Ovchinnikov equations (\ref{El0})-(\ref{El0b}) and (\ref{El2a})-(\ref{El2c}) if we change notations as $\undertilde{f} = -f^+$.

In the diffusive limit, we find the Usadel equation for $\hat G(\mathbf{r},E) = \bigl\langle \hat{g}(\mathbf{r},\hat{\mathbf{k}},E) \bigl\rangle_{\hat{\mathbf{k}}}$, the isotropic part of the $2 \times 2$ Green function:
\begin{align}
& D \nabla_{\mathbf{r}} \left( \hat{G} \nabla_{\mathbf{r}} \hat{G}\right) + i \bigl[ \hat{H_0}, \hat{G} \bigr]=0,
\label{Usadel2x2}
\\
& \hat{H_0}(\mathbf{r},E) =
\begin{pmatrix}
E+i0  &  i \Delta_0(\mathbf{r},E) \\
i \Delta_0^* (\mathbf{r},E) & -E-i0
\end{pmatrix},
\label{H}
\\
& \hat{G}(\mathbf{r},E)
=
\begin{pmatrix}
G & i_s F  \\
i_s \undertilde{F} & - G
\end{pmatrix}_{(\mathbf{r},E)},
\label{g}
\end{align}
which is a reduced form of Eqs.\ (\ref{usadel_app2})-(\ref{UsadelGF}). In Eq.\ (\ref{H}), we have taken into account that only the $s$-wave (hence, singlet) component of the pair potential can survive in the diffusive limit. Therefore we have restored the $\nu=0$ subscript of the $\Delta$ component. This form is also valid in the triplet case, when $\Delta_0 =0$ and Eqs.\ (\ref{Usadel2x2})-(\ref{g}) describe proximity-induced $s$-wave triplet odd-frequency superconducting correlations. The normalization condition takes the form
\begin{equation} \label{norm}
G^2 + i_s^2 F \undertilde{F} = 1.
\end{equation}

\subsection{Single-component pair potential with \texorpdfstring{$\varphi=0$}{varphi=0}}

In the absence of external phase sources (magnetic fields, junctions to superconductors with phase difference), the superconducting phase $\varphi$ is constant, which makes it possible to further simplify equations. Putting the constant phase to zero, we obtain real $\Delta_0$ in the singlet $s$-wave case ($\nu=0$, $s_s=-1$). We can straightforwardly check that the Usadel equation for $\hat G^*(\mathbf{r},-E)$ then has exactly the same form as the original one for $\hat G(\mathbf{r},E)$, see Eq.\ (\ref{Usadel2x2}), which implies the symmetry
\begin{equation} \label{gsymmetry}
\hat G^*(\mathbf{r},-E) = \hat G(\mathbf{r},E).
\end{equation}

At first glance, the same reasoning is valid in the triplet case as well, since this case corresponds to $\Delta_0=0$ (so that $\Delta_0$ can still be considered real). However, this case requires more care. The point is that the Usadel equation with $\Delta_0=0$ does not contain information about the rest of the structure. The superconductivity in the diffusive region is induced from a clean region with nonzero $\Delta_\nu$ (where $\nu=1$, $2$, or $3$), and we must either consider symmetries in the diffusive part taking into account boundary conditions at the interface with the clean region, or choose a more general approach and consider symmetries of the whole structure. We prefer the latter strategy, which requires us to make a step back and consider the Eilenberger-Larkin-Ovchinnikov equation (applicable for the whole structure).

Another complication of the triplet case is that in addition to the overall superconducting phase, there can also be a geometry-dependent internal phase due to anisotropic nature of the parent superconducting state. This is exactly the case for the system we are interested in, where superconductivity originates from a chiral $p$-wave superconductor. An interaction of the form
\begin{equation} \label{Vchiral}
V(\theta,\theta') = V_0 \left( \cos\theta \cos\theta' + \sin\theta \sin\theta' \right)
\end{equation}
leads to the pair potential that can be written as
\begin{equation}
\Delta(\boldsymbol{r},\theta) = e^{i\varphi(\boldsymbol{r})} \left( \Delta_x(\boldsymbol{r}) \cos\theta +i\Delta_y(\boldsymbol{r}) \sin\theta \right),
\end{equation}
with real $\Delta_x$, $\Delta_y$, and $\varphi$. In the absence of external phase sources, the phase $\varphi$ is constant, and we can put it to zero. However, there is also the internal phase of the pair potential (due to $i$ between the two terms), which plays a role depending on the geometry of the problem. Therefore fixing the phase requires fixing the problem geometry.

We are interested in a quasi-one-dimensional problem, when $\Delta$ depends only on $x$ (normal to the surface); then
\begin{equation} \label{DeltaChiral}
\Delta(x,\theta) = \Delta_x(x) \cos\theta +i\Delta_y(x) \sin\theta .
\end{equation}
In this case, we can straightforwardly check that the Eilenberger-Larkin-Ovchinnikov equation for $\hat g^*(x,-\theta,-E)$ has exactly the same form as the original one for $\hat g(x,\theta,E)$, see Eq.\ (\ref{eilenberger_22}), which implies the symmetry
\begin{equation} \label{addsymm}
\hat g^*(x,-\theta,-E) = \hat g(x,\theta,E).
\end{equation}
Averaging over $\theta$ in the diffusive region finally produces the same symmetry relation (\ref{gsymmetry}) as in the singlet case.

As a result, Eq.\ (\ref{gsymmetry}) (valid for both the singlet and triplet states) implies that the case of $\varphi=0$ leads to an additional symmetry,
\begin{equation}
i_s \undertilde{F} (x,E) = i_s^* F (x,E).
\end{equation}
The normalization condition (\ref{norm}) then simplifies to
\begin{equation} \label{normal}
G^2 + F^2 = 1,
\end{equation}
making it possible to employ the standard $\Theta$ parametrization,
$G=\cos\Theta$ and $F=\sin\Theta$. Then we obtain the Usadel equation in the well-known form
\begin{equation} \label{Us_all}
D\nabla^2 \Theta +2i E \sin \Theta + 2 \Delta_0 \cos \Theta =0
\end{equation}
(in the triplet case, this equation can only describe proximity-induced $s$-wave triplet odd-frequency superconductivity in a diffusive region, where $\Delta_0=0$).

Denoting $E_\pm = E\pm \omega/2$, we can write the general expression (\ref{sigma_omega}) for the ac conductivity as
\begin{widetext}
\begin{align}
\frac{\sigma(\mathbf{r},\omega)}{\sigma_0} = \frac 1{2\omega} \int\limits_{-\infty}^\infty dE
\biggl\{
\tanh\Bigl(\frac{E_+}{2T}\Bigr)
& \left[
G^*(E_-) \re G(E_+) + i F^*(E_-) \im F(E_+)
\right]
\notag \\
-\tanh\Bigl(\frac{E_-}{2T}\Bigr)
& \left[
G(E_+) \re G(E_-) - i F(E_+) \im F(E_-)
\right]
\biggr\},
\label{sigma_all}
\end{align}
\end{widetext}
generalizing the expression obtained in Ref.\ \cite{Glazman} to describe not only the singlet but also triplet case.
This is the expression for the complex conductivity employed in the main part of the paper; being represented in terms of the real and imaginary parts, it takes the form of Eqs.\ (\ref{sigma1}) and (\ref{sigma2}) (previously, we used the same expressions in Ref.\ \cite{Asano-Fominov}).

\section{Self-consistency equation}
\label{app:self-cons}

The pair potential and the anomalous part of the Green function are related by the self-consistency equation.
In the case of instantaneous effective electron-electron interaction (which leads to superconductivity), we can write this equation as
\begin{equation}
\Delta_{\alpha\beta}(\mathbf{r},\hat{\mathbf{k}}) = i \pi T \sum_{\omega_n} \bigl\langle V_{\alpha\beta;\gamma\delta}(\hat{\mathbf{k}},\hat{\mathbf{k}}') \mathrm{f}_{\gamma\delta} (\mathbf{r},\hat{\mathbf{k}}',\omega_n) \bigr\rangle_{\hat{\mathbf{k}}'},
\end{equation}
where the Greek indices correspond to the spin space and
the sum runs over Matsubara frequencies with the absolute values smaller than the cutoff set by the Debye frequency.

In the absence of spin-dependent potentials and spin-orbit interaction, the anomalous part of the Green function has the same spin components as the pair potential. Then, in the case of single-component pairing [Eqs.\ (\ref{DeltaSingleComp}) and (\ref{gSingleComp})], the self-consistency equation simplifies as
\begin{equation}
\Delta(\mathbf{r},\theta) = i_s \pi T \sum_{\omega_n} \bigl\langle V(\theta,\theta') f (\mathbf{r},\theta', \omega_n) \bigr\rangle_{\theta'},
\end{equation}
where the scalar quantity $V$ is the interaction in the pairing channel.

In the case of chiral $p$-wave superconductor [see Eq.\ (\ref{Vchiral})] with flat surface, the pair potential can be chosen in the form of Eq.\ (\ref{DeltaChiral}). The real-energy symmetry of Eq.\ (\ref{addsymm}) leads to $f^*(x,-\theta,\omega_n) = - f(x,\theta,\omega_n)$ in the Matsubara representation. Then we obtain the self-consistency equation in the form of Eqs.\ (\ref{Sc0b}) and (\ref{Sc0c}) (which are written in the representation with converging sum, so the summation range can be extended to infinity).


\begin{thebibliography}{999}
\bibitem{comment_fnotations}
Our notations are consistent with the general derivation given in the Appendices. At the same time, the present notations are slightly different from the ones used previously in Refs.\ \cite{Bakurskiy1,Bakurskiy2}. The present form of equations can be reproduced from the old notations of Refs.\ \cite{Bakurskiy1,Bakurskiy2} by substituting $f_\mathrm{old} \mapsto if$ and $f_\mathrm{old}^+ \mapsto -if^+$ (or, equivalently, by substituting $\Delta_\mathrm{old} \mapsto -i \Delta$).
\bibitem{comment:chiralvspy}
Note that a purely $p_y$-wave (or $p_x$-wave) superconducting state is allowed by Eqs.\ (\ref{Sc0b}) and (\ref{Sc0c}) along with the chiral $p_x+ip_y$ state. However, the fully gapped chiral state is more energetically favorable in the bulk than gapless $p_y$ and $p_x$ states (featuring nodal lines).
\bibitem{comment:x=0}
The Green functions in similar structures at $x=0$, i.e., at the interface between the superconductor and the diffusive surface layer, have been calculated in Ref.\ \cite{Bakurskiy1}. While the results are quantitatively different, main qualitative features related to the Andreev band formation are the same as we find in this paper at $x=d$ (free surface of the structure).
\end{thebibliography}
\end{document}